\documentclass[natbib]{svjour3}              
\smartqed  
\usepackage{graphicx}
\usepackage{natbib}
\usepackage{lscape}
\usepackage{longtable}
\newcommand\phs{\phantom{$-$}}
\newcommand\phn{\phantom{0}}
\newcommand\pht{\phantom{00}}
\newcommand\pppp{\phantom{$\pm$}}
\newcommand\aap{A\&A\ }
\newcommand\an{AN\ }
\newcommand\aaps{A\&AS\ }
\newcommand\apss{Ap\&SS\ }
\newcommand\aapr{A\&ARv\ }
\newcommand\apj{ApJ\ }
\newcommand\apjl{ApJL\ }
\newcommand\apjs{ApJS\ }
\newcommand\aj{AJ\ }
\newcommand\mnras{MNRAS\ }
\newcommand\pasp{PASP\ }

\newcommand\nodata{ ~$\cdots$~ }
\journalname{Astronomy \& Astrophysics Review}
\begin{document}

\title{Accurate masses and radii of normal stars: Modern results and applications
}

\titlerunning{Accurate masses and radii of normal stars} 

\author{G. Torres
\and
J. Andersen
\and
A. Gim{\'e}nez
}

\authorrunning{G. Torres et al.} 

\institute{
G. Torres \at
Harvard-Smithsonian Center for Astrophysics; 60 Garden Street, Mail Stop 20, 
Cambridge, MA 02138, USA \\
\email{gtorres@cfa.harvard.edu}           
           \and
J. Andersen \at
The Niels Bohr Institute, Astromomy; University of Copenhagen, Juliane Maries Vej 30, DK - 2100 Copenhagen, Denmark \\
{\it and} Nordic Optical Telescope Scientific Association, La Palma, Spain\\
\email{ja@astro.ku.dk}           
           \and
A. Gim{\'e}nez \at 
Centro de Astrobiolog\'{i}a (CSIC/INTA); Carretera de Torrej\'on a Ajalvir, km 4,
E - 28850 Torrej\'on de Ardoz, Madrid, Spain\\
\email{agimenez@rssd.esa.int}           
}

\date{Received: date / Accepted: date}

\maketitle

\begin{abstract}

This paper presents and discusses a critical compilation of accurate, fundamental determinations of stellar masses and radii. We have identified 95 detached binary systems containing 190 stars (94 eclipsing systems, and $\alpha$~Centauri) that satisfy our criterion that the mass and radius of both stars be known to $\pm$3\% or better. All are non-interacting systems, so the stars should have evolved as if they were single.  This sample more than doubles that of the earlier similar review by \cite{A91}, extends the mass range at both ends and, for the first time, includes an extragalactic binary. In every case, we have examined the original data and recomputed the stellar parameters with a consistent set of assumptions and physical constants. To these we add interstellar reddening, effective temperature, metal abundance, rotational velocity and apsidal motion determinations when available, and we compute a number of other physical parameters, notably luminosity and distance.

These accurate physical parameters reveal the effects of stellar evolution with unprecedented clarity, and we discuss the use of the data in observational tests of stellar evolution models in some detail. Earlier findings of significant structural differences between moderately fast-rotating, mildly active stars and single stars, ascribed to the presence of strong magnetic and spot activity, are confirmed beyond doubt. We also show how the best data can be used to test prescriptions for the subtle interplay between convection, diffusion, and other non-classical effects in stellar models.

The amount and quality of the data also allow us to analyse the tidal evolution of the systems in considerable depth, testing prescriptions of rotational synchronisation and orbital circularisation in greater detail than possible before. We show that the formulae for pseudo-synchronisation of stars in eccentric orbits predict the observed rotations quite well, except for very young and/or widely-separated stars. Deviations do occur, however, especially for stars with convective envelopes. The superior data set finally demonstrates that apsidal motion rates as predicted from General Relativity plus tidal theory are in good agreement with the best observational data. No reliable binary data exist that challenge General Relativity to any significant extent.

The new data also enable us to derive empirical calibrations of $M$ and $R$ for single (post-) main-sequence stars above 0.6~$M_{\odot}$. Simple, polynomial functions of $T_{\rm eff}$, $\log g$ and [Fe/H] yield $M$ and $R$ with errors of 6\% and 3\%, respectively. Excellent agreement is found with independent determinations for host stars of transiting extrasolar planets, and good agreement with determinations of $M$ and $R$ from stellar models as constrained by trigonometric parallaxes and spectroscopic values of $T_{\rm eff}$ and [Fe/H]. 

Finally, we list a set of 23 interferometric binaries with masses known to better than 3\%, but without fundamental radius determinations (except $\alpha$ Aur). We discuss the prospects for improving these and other stellar parameters in the near future.

\keywords{Stars: fundamental parameters \and Stars: binaries: eclipsing \and Stars: binaries: spectroscopic \and Stars: interiors \and Stars: evolution}

\end{abstract}

\section{Introduction}\label{intro}

Stars are the main baryonic building blocks of galaxies and the central engines in their evolution. Accurate knowledge of the masses, radii, luminosities, chemical compositions, and ages of stars is fundamental to our understanding of their structure and evolution. This, in turn, underlies our models of the nucleosynthesis in stars and their interaction with their environment -- the driving forces in the evolution of galaxies. It is important, therefore, to establish the basic properties of stars using a minimum of theoretical assumptions -- preferably only geometry and Newtonian mechanics.

The required accuracy of these fundamental data depends on the intended application. When estimating the bulk properties of large groups of stars, an uncertainty of 10\% may be quite acceptable, while for other purposes such data are useless. Testing models of stellar evolution is perhaps the most demanding application: The sensitivity of all properties of a stellar model to the initial mass is such that virtually any set of models will fit any observed star, if the mass is uncertain by just $\pm$5\%. Only data with errors below $\sim$1--3\% provide sufficiently strong constraints that models with inadequate physics can be rejected. 

The aim of this paper is to present a critical assessment and summary of the currently available fundamental determinations of stellar masses and radii of sufficient accuracy for even the most demanding applications. Thus, this paper supersedes the earlier review by \cite{A91} (A91). As before, the stars included in our sample are all members of detached, non-interacting binary systems: Only for stars with detectable companions can the mass be determined directly and with errors of $\sim$1\%, and only in eclipsing binary systems can both the stellar masses {\it and} radii be determined to this accuracy (with the sole exception of $\alpha$ Centauri). We identify 95 binary systems in the literature that satisfy these selection criteria. 

Significant progress since the earlier review includes the large body of data published since 1991, reflecting the improvements in observing and analysis techniques. In this paper, we also systematically recompute the stellar masses and radii from the original data with the same analysis techniques and modern, consistent values for the associated physical constants. Similarly, we critically review the published effective temperatures of the stars, which are used to compute their radiative properties. The stellar parameters listed in Table~\ref{tableMR} are therefore not necessarily exactly identical to those given in the original papers.

To facilitate further discussion of the data presented here, we also provide individual rotational velocities, reddenings, metal abundances and distances whenever possible, and we compute approximate ages for all the systems. Finally, we list relevant additional data for the 29 systems that have well-studied apsidal motion.

Following a brief overview of the data, we discuss their use in testing state-of-the-art stellar evolution models. We also discuss the effects of tidal evolution in these systems, now from much-improved observational data, and explore the predictions of tidal theory for the axial rotation of binary components and the Newtonian and general relativistic contributions to the apsidal motion of the eccentric systems. 

The data also allow us to devise calibrations that provide good estimates of mass and radius for single stars with reliable determinations of $T_{\rm eff}$, $\log g$ and [Fe/H]. 

Progress in long-baseline optical interferometry offers much promise for the future determination of stellar masses and radii, although the radii have not yet reached the level of accuracy aimed at here. We therefore list a number of (non-eclipsing) interferometric binaries in which the mass errors meet our selection criterion, but where more work is needed for the radii to also do so.

The paper ends with a number of considerations for the future.

We emphasise here that the criterion for inclusion in this review is quality of the data only, not any attempt to make the sample complete or unbiased in any sense. We warn the reader, therefore, that it is {\it unsuitable for any kind of statistical analysis}, its other qualities notwithstanding.

\section{Selection criteria}\label{selection}

The basic criteria for selecting binary systems for this review are, first, that the components can be expected to have evolved as if they were single, second, that their masses and radii can both be trusted to be accurate to better than 3\%. We have endeavoured to make an exhaustive search of the literature for such systems.

The first requirement excludes all systems with past or ongoing mass exchange, and systems exhibiting an activity level far exceeding that seen in single stars, such as cataclysmic, Algol, or RS~CVn-type binaries. The second condition obliges us to perform an in-depth study of the pedigree of each eclipsing system, following the precepts outlined in Sect.~\ref{analysis}.

The choice of the upper limit of permissible errors is a matter of judgement. The nominal limit in the compilation in A91 was 2\% in both $M$ and $R$, but several systems did in fact have errors between 2\% and 3\%. We have adopted 3\% as a hard cut-off for the error in both parameters for the systems included in Table~\ref{tableMR}, recognising that an error of 3\% may be too large for firm conclusions in, e.g., the most demanding tests of stellar evolution theory (see Sect.~\ref{sysfit}). However, if needed, readers will be able to select subsamples with stricter limits from the data given here.

Due to the advances in long-baseline optical/near-IR interferometry, a sizeable number of visual binaries now have combined interferometric and spectroscopic orbits yielding individual masses with errors below 3\%. The orbital parallaxes and resulting absolute magnitudes are also of high accuracy. However, the individual radii of these stars are not accurate enough to be included in Table~\ref{tableMR}, whether determined from direct angular diameter measurements (as for Capella) or from the absolute magnitudes and effective temperatures, as the error of $T_{\rm eff}$ is still too large.

Nevertheless, these systems contain a number of interesting objects with very well-determined masses, and prospects are good that progress in interferometry over the next few years may improve the accuracy of the radii to match that of the masses. We therefore list and briefly discuss these systems in Table~\ref{tableMonly} and Sect.~\ref{Mdata}.

\section{Analysis techniques}\label{analysis}

Results of the highest accuracy require complete, high-quality data, analysed with appropriate techniques and with a critical assessment of formal and -- especially -- systematic errors. Criteria for and examples of suitable observational data were discussed in A91 (chiefly, accuracy and phase coverage of the light and radial-velocity curves). We have inspected all the original data used in the determinations listed in Table~\ref{tableMR} to satisfy ourselves that they are adequate to support the published accuracies. 

A detailed discussion of state-of-the-art analysis techniques as of 1991 for both spectroscopic and photometric data was given in A91. This need not be repeated here, but in the following we briefly review the main developments in observational and computational techniques since that time.

\paragraph{Mass determination.\label{massdet}} 
The most critical requirement for obtaining accurate masses is an accurate determination of the orbital velocities from the observed double-lined spectra, both for eclipsing and non-eclipsing binaries, because the derived masses are proportional to the third power of these velocities. This requires spectra of good spectral resolution and S/N ratio, properly analysed.

The most significant progress in the intervening two decades is the great advance in digital spectroscopy, chiefly through the use of modern \'echelle spectrographs and CCD detectors, coupled with the perhaps even greater advances in numerical analysis techniques. Today's binary star observers employ spectra of much higher resolution and S/N ratio than the vast majority of the studies reported in A91, and accurate velocities can be derived with sophisticated mathematical techniques. These include two-dimensional cross-correlation techniques (\citealt{todcor}, extended to systems with three or four components by \citealt{tricor, quadcor}), the broadening function technique \citep{1992AJ....104.1968R}, and several variants of the `disentangling' technique \citep[e.g.,][]{1991ApJ...376..266B, simon, hadrava, gonzalez} operating either in wavelength space or Fourier space. 

The disentangling method takes advantage of the fact that a set of spectra distributed over the orbital cycle of a double-lined binary displays the same two spectra, only shifted by different relative velocities. Best-estimate values for the two spectra and the orbital elements are then extracted from the observations by a statistical analysis technique. The individual spectra can then be further analysed by standard single-star procedures to derive effective temperatures and chemical compositions for the two stars -- a significant additional advantage. The determination of individual radial velocities is optional in this technique, since the orbital elements can be fit directly to the spectra with no need for an intermediate stage of measuring actual Doppler shifts \citep[see, e.g.,][]{1998A+A...331..167H}. Similarly, the orbital elements can be fit directly to an ensemble of one- or two-dimensional 
cross-correlation functions, as done, e.g., by \cite{1997ApJ...485..167T} and \cite{1999A+A...351..619F}. A minor inconvenience is that the orbital fit cannot easily be visualized, since there are no velocities to display.

The critical point, in line-by-line measurements as well as in the above more powerful, but less transparent methods, is to ascertain that the resulting velocities and orbital elements are free of systematic error. The most straightforward test today is to generate a set of synthetic binary spectra from the two component spectra and the orbital elements as determined from a preliminary analysis, computed for the observed phases and with a realistic amount of noise added. The synthetic data set is then analysed exactly as the real data, and the input and output parameters are compared. The differences, if significant, are a measure of the systematic errors of the procedure, and can be added to the real observations to correct for them \citep[see, e.g.,][]{popper94, dmvir, hshya}.

The least-squares determination of spectroscopic orbital elements from the observed radial velocities is in principle straightforward. However, subtle differences exist between various implementations, which may lead to somewhat different results from the same data sets. We have therefore systematically recomputed orbital elements from the original observations. 

A point of minor importance, but still significant in mass determinations of the highest accuracy possible today, is the value of the physical constants used to compute the stellar masses and orbital semi-axis major from the observed orbital parameters. The recommended formulae are:
\begin{eqnarray*}
M_{1,2} \sin^3 i &=& 1.036149\times10^{-7} (1-e^2)^{3/2} (K_1 + K_2)^2 K_{2,1} P \\
a \sin i &=& 1.976682\times10^{-2} (1-e^2)^{1/2} (K_1 + K_2)P
\end{eqnarray*}
where $M_{1,2}\sin^3i$ are in units of solar masses, the orbital velocity amplitudes $K_{2,1}$ are in km s$^{-1}$, the orbital period $P$ in days, and the orbital semi-axis major in solar radii; $i$ and $e$ are the orbital inclination and eccentricity, respectively. The numerical constants given above correspond to the currently accepted values, in SI units, for the heliocentric gravitational constant, $GM_{\odot} = 1.3271244 \times 10^{20}$~m$^3$~s$^{-2}$ \citep[see][]{standish} and the solar radius, $R_{\odot} = 6.9566 \times 10^8 $~m \citep{2008ApJ...675L..53H}. However, some authors still use the old value of $1.0385\times10^{-7}$ for the constant in the mass formula -- a difference of 0.23\%, which is not entirely negligible by today's standards. Uncertainties in the solar radius itself also affect the stellar radii when expressed in that unit, but at a level $< 0.1$\%, at which the very definition of the radius of a star comes into play.

Accordingly, we have recomputed all the masses and radii listed in Table~\ref{tableMR} from our own solutions of the original observations, using the constants listed above.

\paragraph{Light curve analysis.\label{lcanalysis}} 
A variety of codes exists to analyse the light curves of eclipsing binaries and derive the orbital parameters ($i$, $e$, and $\omega$) and stellar radii in units of the orbital semi-axis major. The most frequent obstacle to an accurate radius determination from such codes, notably in partially eclipsing systems, is the fact that a wide range of combinations of stellar radii, $i$, $e$, and $\omega$, may yield light curves that are essentially indistinguishable. 

Whether or not convergence is easy, the results must therefore always be checked against the spectroscopic determination of $e$ and $\omega$ and the temperature and luminosity ratio of the two stars. Because the luminosity ratio is proportional to the square of the ratio of the radii, it is a particularly sensitive -- often indispensable -- constraint on the latter \cite[see][for an extreme example]{tzfor}. 

The relative depths of the light curve minima and the colour changes during eclipse are usually robustly determined from the light curve solution and are good indicators of the surface flux ratio between the two stars. Whenever possible, we have used these data to check the published temperature differences between the two stars. 

Formal error estimates from the codes rarely include the contribution of systematic effects. Comparing separate solutions of light curves in several passbands is one way to assess the reliability of the results; computing light curves for several parameter combinations and evaluating the quality of the fit to the data is another. 

\paragraph{Consistency checks.\label{consistency}}
In all cases, it is important to verify the consistency of the different types of information on a given system as thoroughly as possible. The values for the period (rarely a problem except in systems with apsidal motion), $e$ and $\omega$ of the orbit must be internally consistent, as must the luminosity ratio of the components as measured from the light curves and seen in the spectra. 

Certain light-curve codes, e.g., the widely-used WD code \citep{WD}, allow one to input a set of light curves in several colours as well as the radial-velocity observations, and return a single set of results for the stellar and orbital parameters. From a physical point of view this is clearly the preferable procedure, and the resulting single sets of masses, radii, etc.\ often have impressively small formal errors. 

However, such a procedure tends to obscure the effects of flaws and inconsistencies in the data and/or the binary model, and consistency checks such as those described above become difficult or impossible. If consistency has been verified independently, a combined, definitive solution can be performed with confidence, but the basic philosophy should always be that consistency is a condition to be verified, not assumed.

In compiling the data presented in Table~\ref{tableMR}, we have verified that the conditions described in this section are satisfied in every system, usually by recomputing the stellar parameters from the original data. The numbers presented here will therefore not always be strictly identical to those found in the original analyses.

\section{Additional data}\label{auxdata}

Mass and radius are the data that can be determined directly from observation without relying on external data or calibrations. However, to fully utilise the power of these parameters, additional data are needed. These are, most importantly, the effective temperature and chemical composition of the stars, followed by their rotational velocities and the amount of interstellar reddening; the latter is needed when deriving effective temperatures, luminosities, and distances. We provide these data for the systems in Table~\ref{tableMR} as far as possible, and briefly describe our selection of them here.

\paragraph{Effective temperature.\label{teff}}
Effective temperatures are usually determined from multicolour photometry via an appropriate calibration, although spectroscopic excitation temperatures are used occasionally when the two spectra can be separated. The determinations available in the literature are rather heterogeneous, being typically based on photometry in a variety of systems as selected by the original observers, and using a variety of calibrations as necessary to cover a temperature range from 3,100 K to 38,000 K. 

It is an essentially impossible task to place all the temperature determinations on a consistent -- let alone correct -- scale; this would require obtaining new optical and IR photometry for many systems and an in-depth review of the corresponding temperature calibrations, a task well beyond the scope of this paper. Instead, we have checked the data, calibrations, and determinations of interstellar reddening, if any, in the original papers, and have searched the literature for any other reddening determinations. When known, our adopted $E(B-V)$ value for each system is listed in Table~\ref{tableMRsup} and has been used to estimate $T_{\rm eff}$ and its uncertainty. The resulting values of $T_{\rm eff}$ have typical errors of $\sim$2\%, but some stars have considerably larger errors, as indicated in Table~\ref{tableMR}.

From the measured radius $R$ and the adopted value of $T_{\rm eff}$, the luminosity $L$ of each star is computed and listed in Table~\ref{tableMR}. Adopting the scale of bolometric corrections $BC_V$ by \cite{flower} and a consistent value of $M_{\rm bol,\odot}$ (essential!), the absolute visual magnitude $M_V$ follows and is reported as well; note that the \cite{flower} $BC_V$ may not be accurate for the lowest-mass stars. 

\paragraph{Distances.\label{distance}}
From the apparent visual magnitudes $V_{\rm max}$ listed in Table~\ref{tableMR}, $M_V$ as computed above, and the reddenings listed in Table~\ref{tableMRsup}, the distance of each system follows and is also given in Table~\ref{tableMRsup}. As seen, the accuracy of these distances is remarkably good, with an average error of only 5\% (disregarding a few outliers). This has led to the use of extragalactic eclipsing binaries as distance indicators -- including one system listed here (OGLE~051019). It must be recalled, however, that these distances are sensitive to any systematic errors in $T_{\rm eff}$, on which they depend as $T_{\rm eff}^2$. 

\begin{figure}[t]
\center \includegraphics[width=0.7\textwidth]{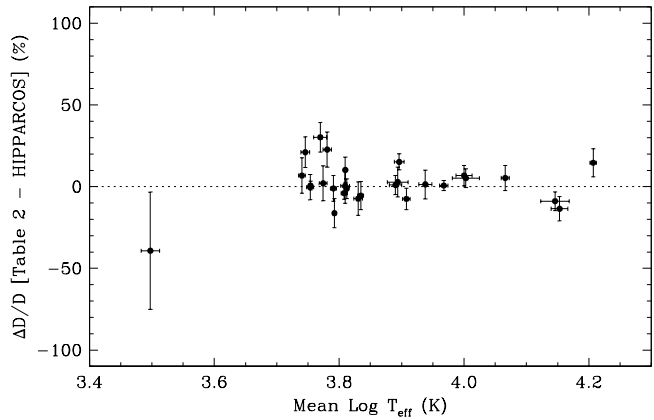}
\caption{Percentage difference between the distances $D$ in Table~\ref{tableMR}
and as derived from the revised Hipparcos parallaxes (both with errors $<$10\%), as a function of mean system $\log T_{\rm eff}$.}
\label{distcheck}
\end{figure}

Fortunately, an external check of the distances (and hence of $T_{\rm eff}$) is provided by the Hipparcos parallaxes \citep{Perryman:97}, revised recently by \cite{vanleeuwen}. There are 28 systems for which the distances derived in Table~\ref{tableMRsup} and from the revised Hipparcos parallaxes are both known to better than 10\%. Fig.~\ref{distcheck} shows the difference (in percent) between the two determinations, as a function of $T_{\rm eff}$. Weighting these differences according to their errors, the mean difference and standard deviation are 1.5\% and 8.2\%, respectively, and no systematic trend is apparent. This comparison suggests that the adopted effective temperature scales are essentially correct and adds strong support to the use of binary stars as distance indicators. However, close attention to the determination of $T_{\rm eff}$ and interstellar reddening is needed in every case. 

\paragraph{Metal abundances.\label{metallicity}}
Knowledge of the chemical composition of a star is needed in order to compute appropriate theoretical models for its evolution. Accordingly, we have searched the literature for [Fe/H] determinations for the stars in Table~\ref{tableMR}; the 21 reliable values found are reported in Table~\ref{tableMRsup}. Photometric metallicity determinations are subject to a number of uncertainties, including interstellar reddening, so we have elected to include only spectroscopic determinations here. In most cases, they refer to the binary itself, but we have included a few systems in open clusters with [Fe/H] determinations from other cluster members. 

Most of the systems in Table~\ref{tableMRsup} have metallicities close to solar, so a solar abundance pattern for the individual elements is expected \citep[see, e.g.,][]{edv93}. Hence, the $Z$ parameter of stellar models should generally scale as [Fe/H].

\paragraph{Rotational velocities.\label{rotation}}
Axial rotational speed is an important input parameter in light curve synthesis codes as it enters the computation of the shapes of the stars. In addition, the axial rotations and their degree of synchronisation with the orbital motion are direct probes of the tidal forces acting between the two stars \citep[e.g.,][]{mazeh}, as is the orbital eccentricity. Rotations are also needed to compute apsidal motion parameters for binaries with eccentric orbits (see Sect.~\ref{apsidal}). 

We have therefore collected the available direct spectroscopic determinations of $v\sin i$ and its uncertainty for as many of the components of the stars in Table~\ref{tableMR} as possible (81 systems); the results are reported in Table~\ref{tableMRsup}. While the accuracy varies from system to system, it is usually sufficient to detect appreciable deviations from the default assumption of (pseudo-)synchronism. 

\paragraph{Ages.\label{ages}}
The age of a star can, in principle, be determined from theoretical isochrones when $T_{\rm eff}$, $\log g$ (or $M_V$), and [Fe/H] are known. Computing stellar ages and --   especially -- their uncertainty in practice is, however, a complex procedure fraught with pitfalls \citep[see, e.g., the extensive discussion in][]{GCS1}. Some of the original sources of the data in Table~\ref{tableMR} discuss ages, but many are based on outdated models, and their origin is necessarily heterogeneous. Determining truly reliable ages for the stars in Table~\ref{tableMR} would require (re)determination of effective temperature and [Fe/H] for many systems and a detailed comparison with models (see Sect.~\ref{models}) -- a major undertaking, which is well beyond the scope of this review.

Even a rough age estimate is, however, a useful guide to the nature of a system under discussion, e.g., when assessing the degree of circularisation of the orbit and/or synchronisation of the rotation of the stars. To provide such estimates on a systematic basis, we have computed ages for all but the lowest-mass systems from the Padova isochrones \citep{girardi}, taking our adopted values of $T_{\rm eff}$, $\log g$, and [Fe/H] as input parameters and setting [Fe/H] = 0.00 (solar) when unknown. For the lowest-mass systems and some of the more recent studies, we have adopted the ages reported in the original papers. All the age estimates are listed in Table~\ref{tableMRsup}.

{\it We emphasise that these values are indicative only:} Errors of 25--50\% are likely, and in several cases they will be larger. Any accurate age determination and realistic assessment of its errors would require a critical re-evaluation of the input parameters $T_{\rm eff}$ and [Fe/H] -- probably requiring new observations -- and far more sophisticated computational techniques than are justified with the material at hand. For this reason, we deliberately do not give individual error estimates for the ages in Table~\ref{tableMRsup}.

\section{Stars with accurate masses and radii}\label{MandRdata}

The basic and derived parameters for the 95 systems that satisfy all our selection criteria are listed in Tables~\ref{tableMR} and \ref{tableMRsup}. This is more than double the number of systems in A91 and includes improved results for several of the systems listed there. In addition, Table~\ref{tableMRsup} now also provides reddening, distance, $v \sin i$, [Fe/H], and age whenever available, with references to the data for each system. In this section we illustrate various relations between the data and comment briefly on each diagram.

\begin{figure}[ht]
\center \includegraphics[width=0.7\textwidth]{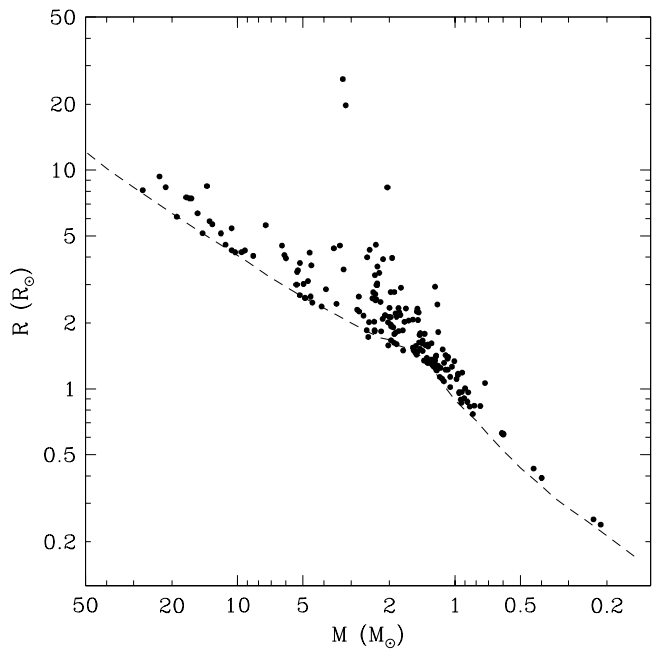}
\caption{$R$ vs.\ $M$ for the stars in Table~\ref{tableMR}; error bars are smaller than the plotted symbols. A theoretical zero-age main sequence (ZAMS) for solar metallicity from \cite{girardi} is shown by the dashed line.}
\label{logRlogM}
\end{figure}

\paragraph{The mass-radius diagram.}
Fig.~\ref{logRlogM} shows the mass-radius diagram for the stars in Table~\ref{tableMR}; note that the error bars are smaller than the plotted symbols. Relative to A91, the mass range has been expanded both at the higher (V3903~Sgr) and the lower end (CM~Dra), and the diagram is, of course, much better populated than before. For the first time it includes an extragalactic binary, the two-giant system OGLE-051019.64-685812.3 in the Large Magellanic Cloud (here called OGLE~051019 for short). It is noteworthy, however, that the two stars in OGLE~051019 are the only new {\it bona fide} red giants since the two in A91 (TZ~For A and AI~Phe A). The large range in radius for a given mass clearly shows the effect of stellar evolution up through the main-sequence band, which in this diagram moves a star up along a vertical line as it evolves, if no significant mass loss occurs.

\begin{figure}[ht]
\center \includegraphics[width=0.8\textwidth]{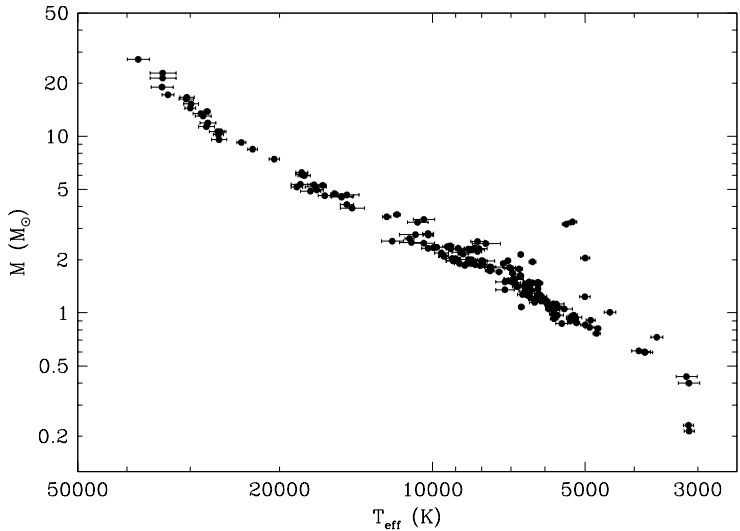}
\caption{$M$ vs.\ $T_{\rm eff}$ for the stars in Table~\ref{tableMR}.}
\label{logMlogT}
\end{figure}

\paragraph{The temperature-mass-radius diagrams.}
Fig.~\ref{logMlogT} shows the observed $M$ vs.\ $T_{\rm eff}$ for the stars in Table~\ref{tableMR}, corresponding roughly to the dependence of mass on spectral type. The effects of stellar evolution are again clearly seen, with stars moving horizontally to the right towards cooler temperatures as they evolve (assuming no mass loss). Note again that only the errors in $T_{\rm eff}$ are large enough to be visible, while main-sequence masses for a given $T_{\rm eff}$ may vary by 40\% or more.

\begin{figure}[ht]
\center \includegraphics[width=0.8\textwidth]{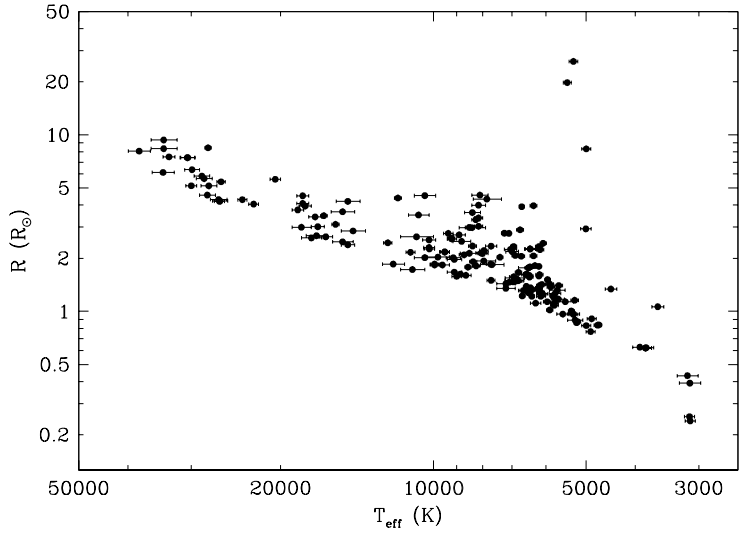}
\caption{$R$ vs.\ $T_{\rm eff}$ for the stars in Table~\ref{tableMR}.}
\label{logRlogT}
\end{figure}

The evolutionary changes are seen even more clearly in Fig.~\ref{logRlogT}, equivalent to a plot of radius vs.\ spectral type. Here, however, as both temperature and radius change during the evolution, the stars will move roughly diagonally towards the upper right. The range in $R$ for a given $T_{\rm eff}$ is much larger than for $M$, up to a factor of $\sim$3 -- again far more than the tiny errors in the individual values of $R$. 

\begin{figure}[ht]
\center \includegraphics[width=0.8\textwidth]{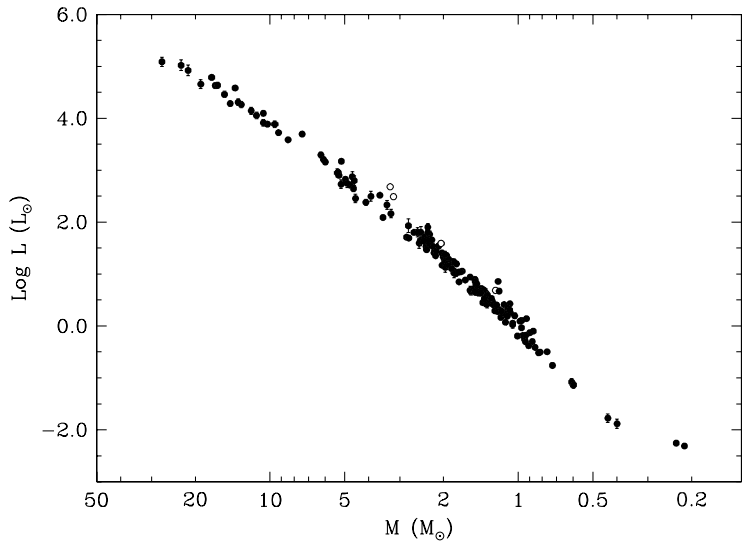}
\caption{The mass-luminosity relation for the stars in Table~\ref{tableMR}. Error bars are shown, and stars classified as giants are identified by open circles. See Sect.~\ref{models} for a discussion of the effects of evolution in this diagram.}
\label{logLlogM}
\end{figure}

\paragraph{The mass-luminosity diagram.}
Fig.~\ref{logLlogM} shows the mass-luminosity relation, i.e.\ the computed 
$\log L$ vs.\ the observed mass, for the stars in Table~\ref{tableMR}. This
relation is popular in a number of modelling contexts and appears very 
well-defined here at first sight. Note, however, that due to the high accuracy of the masses and large range in $\log L$, the error bars are smaller than the plotted symbols. Thus, the scatter seen is of astrophysical, not observational origin, and is due to the varying effects of stellar evolution and chemical abundance from star to star. We elaborate further on these issues in the following section.

\section{Testing stellar models}\label{models}

Comparison with stellar evolution models is one of the most prominent uses of accurate binary data \citep[see, e.g.,][]{Pols:97, Lastennet:02, Hillenbrand:04} and features in most modern papers on binary system parameters. An extensive discussion on the subject was given in A91, with a focus on what information can be obtained from data of increasing degrees of completeness. Only a few main points will be repeated here, with a summary of recent developments.

The key point of the A91 discussion was that, while even the best data can never {\it prove} a set of models right, sufficiently accurate and complete data can reveal significant deficiencies in the physical descriptions in stellar models: When the (preferably unequal) masses and identical composition of two binary components are known, and the age of the model of one star is fixed from its radius, requiring the model to match the observed radius of the other star at the same age is a non-trivial test. Matching the observed temperatures as well provides additional constraints, e.g., on the helium content and mixing length parameter of the models. AI Phe, as shown in A91, probably remains the best textbook example (but see Sect. \ref{sysfit}).

In the following, we briefly review the ways that accurate and increasingly complete binary data can be used to constrain stellar model properties. We begin with some general considerations of the possible tests, then discuss fits to individual systems, and finally review recent progress of more general interest.

\subsection{General considerations}

\paragraph{Information from $M$ and $R$ only.}
The change in radius with evolution through the main-sequence band is clearly seen in the mass-radius diagram of Fig.~\ref{logRlogM}. In the absence of significant mass loss, evolution proceeds vertically upwards in this diagram, and a line connecting the two stars in a given binary system indicates the slope of the isochrone for the age of the system.

The level of the ZAMS (zero-age main sequence) and the slope of young isochrones in this diagram depend on the assumed heavy-element abundance $Z$ of the models, the ZAMS models having larger radii at higher $Z$. Thus, any point below the theoretical ZAMS curve in Fig.~\ref{logRlogM} would be interpreted as that star having a lower metallicity than that of the models, and a binary with a mass ratio sufficiently different from unity can constrain the range in $Z$ of acceptable models. Obviously, the smaller the observational errors, the stronger the constraints on the models \citep[see, e.g.,][]{uoph}. But unless $Z$ (i.e.\ [Fe/H]) has actually been observed, one cannot check whether the model $Z$, and hence the derived age, is in fact correct. And there is still no constraint on the helium abundance and mixing length parameter of the models.

\paragraph{Information obtainable with $M$, $R$, and $T_{\rm eff}$.}
Adding $T_{\rm eff}$ to the known parameters allows one to go a step further. E.g., in the temperature-radius diagram (Fig.~\ref{logRlogT}), models for the accurately known mass of every star have to match not only the two observed {\it radii} for the components of each binary system at the same age, but also the two values of $T_{\rm eff}$. A match can often be obtained by adjusting the metal and/or helium abundance (equal for the two stars) and/or the mixing-length parameter (also the same unless the stars have very different structures), and plausible numbers will usually result, but the {\it test} is weak without a reality check on these numbers.

\paragraph{Tests using $M$, $R$, $T_{\rm eff}$, and {\rm [Fe/H]}.}
Having the complete set of observable data allows one to make truly critical tests of a set of stellar models: $M$ and [Fe/H] fix the basic parameters of the model of each star (assuming a value for the helium abundance $Y$, which normally cannot be observed directly, and/or for the mixing-length parameter). A model for the stars in a binary is then only successful if the temperatures and radii (or luminosities) of both stars are fit within the observational errors at a single age -- a non-trivial requirement.

\begin{figure}[ht]
\center \includegraphics[width=0.7\textwidth]{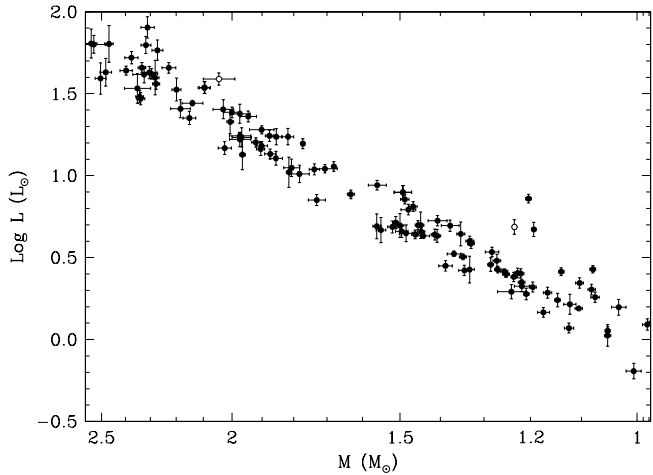}
\caption{Close-up of the 1--2.5 $M_{\odot}$ range of the mass-luminosity relation in Fig.~\ref{logLlogM}. The very significant scatter in $\log L$ at each mass value is due to the combined effects of stellar evolution and abundance differences (see text). Open circles: stars classified as giants.}
\label{logLlogMz}
\end{figure}

The significant effects of both stellar evolution and abundance differences are well seen in the close-up of the deceptively tight mass-luminosity relation of Fig.~\ref{logLlogM} that we show in Fig.~\ref{logLlogMz}. Making the error bars visible highlights the fact that the scatter is highly significant and not due to observational uncertainties. The open symbols show -- not surprisingly -- that the stars classified as giants are more luminous than main-sequence stars of the same mass, but the more subtle effects of evolution through the 
main-sequence phase are also clearly seen.

But evolution is not all, as seen by comparing either star in VV~Pyx with the primary of KW~Hya (nos.\ 42 and 46 in Table~\ref{tableMR}). The stars are virtually identical in mass and radius (or $\log g$), hence in very similar stages of evolution, but their temperatures are quite different and the luminosities differ by nearly a factor of two. Clearly, the two systems are expected to have different compositions, with KW~Hya likely more metal-rich than VV~Pyx. Unfortunately, no actual determination of [Fe/H] is yet available for either system to test this prediction.

\subsection{Fitting individual systems}\label{sysfit}

The most informative comparison of stellar models with real stars is obtained when the mass, radius, temperature, and [Fe/H] are accurately known for both stars in a binary system. If the stars differ significantly in mass and degree of evolution, fitting both stars simultaneously for a single age provides a very stringent test of the models. We have calculated individual evolutionary tracks for the observed masses and metallicities of the systems in Table~\ref{tableMR}, setting [Fe/H] = 0.00 if the metallicity is unknown. In most cases, a respectable fit is achieved, and any modest deviations can usually be explained in terms of uncertain temperatures, reddening and/or metallicity. Resolving the exceptional cases of large unexplained discrepancies will require detailed studies, perhaps involving additional observations, which are beyond the scope of this paper.

\begin{figure}[ht]
\center \includegraphics[width=0.8\textwidth]{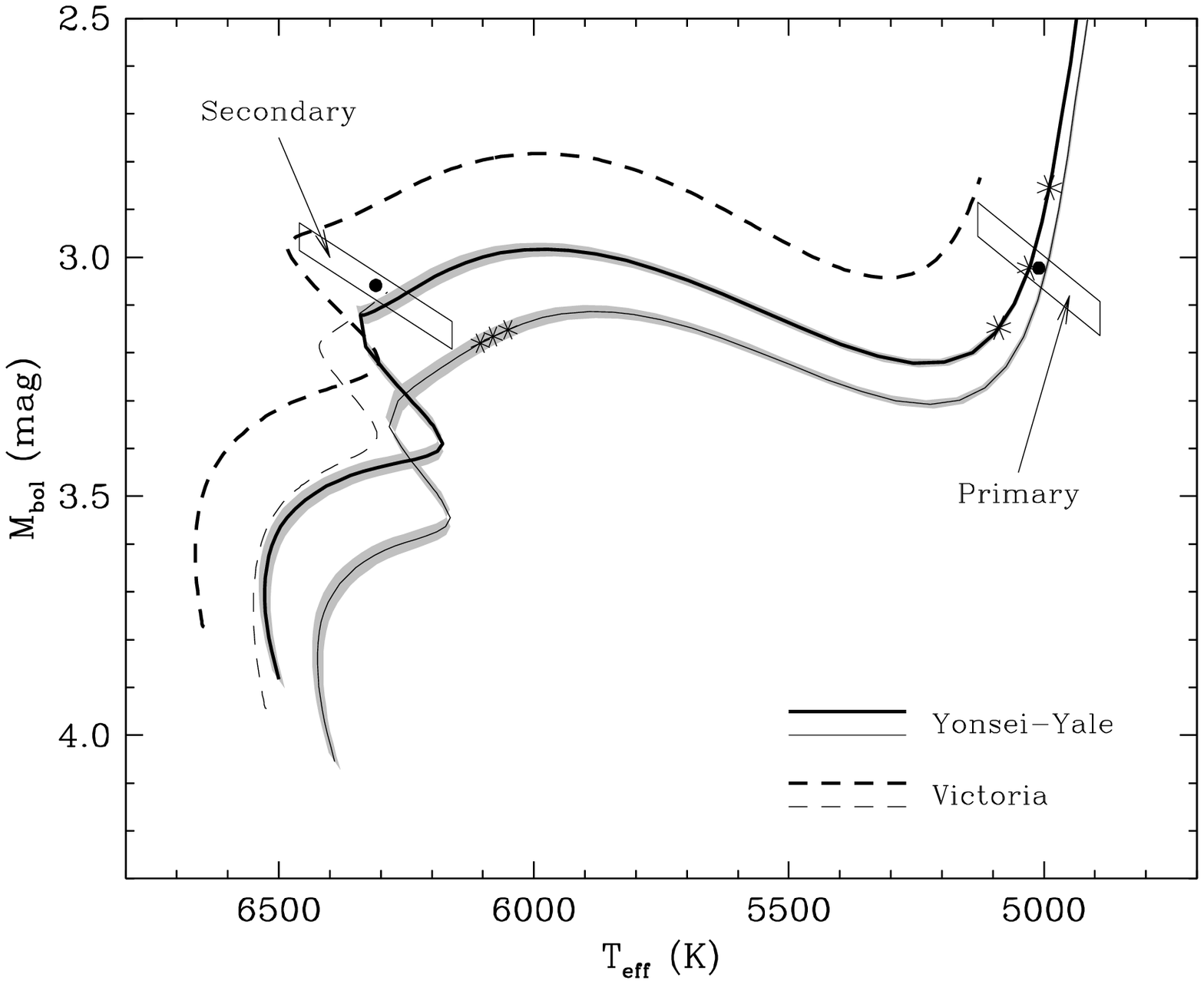}
\caption{Detailed comparison of the observed properties of AI~Phe \citep{aiphe} with models for the observed masses from the Yonsei-Yale \citep[solid lines;][]{yi, demarque} and Victoria codes (VandenBerg, priv. comm.; dashed). The latter extend only until they reach the measured values. Shaded areas indicate the uncertainty in the track locations due to the 0.4\% error in the masses. Models for the primary are drawn with heavier lines. Asterisks on the Yonsei-Yale primary track indicate the best age of that star and changes of $\pm$1\%, and are repeated on the secondary track to illustrate the different speed of evolution.}
\label{aiphe}
\end{figure}

In nearly equal-mass binaries, the requirement for consistency between the two components is only a weak constraint on the models, at best. But the rare examples of significant differential evolution can be very informative, as shown in the classic case of AI~Phe \citep{aiphe} -- see Fig.~\ref{aiphe}. With masses only 20\% larger than that of the Sun and a metallicity only slightly lower, solar calibrations were adopted in that work for the helium content and mixing length of the Victoria models of the time, and a picture-perfect fit was obtained for both stars at exactly the same age.

In order to see how modern stellar evolution codes fare in this comparison, we show in Fig.~\ref{aiphe} the observed properties of AI~Phe together with tracks from the Yonsei-Yale code \citep{yi, demarque} for the measured masses and metallicity (solid lines). As seen, these models fit the primary (cooler) star well, but the track for the secondary (lower curve) is just outside the $1\sigma$ error limit of the observations. Asterisks indicating $\pm$1\% age differences show just how sensitive the fit is. 

At our request, Dr.\ D.\ A.\ VandenBerg kindly computed new models for AI~Phe with an experimental version of the Victoria code \citep{davb06}, which includes He diffusion in the outer layers; note that the adopted mixing length and overshooting parameters of these models have not yet been adjusted to match the solar and other constraints satisfied by the \cite{davb06} model series.

These tracks are also shown in Fig.~\ref{aiphe} (dashed lines) and now match the secondary very well (lower curve), but not the primary -- as expected for a non-optimised mixing length parameter. The derived ages differ by 0.08~Gyr (1.6\%) for the Yonsei-Yale models, and by 0.20~Gyr (5\%) for the experimental Victoria models (cf.\ the 1\% age differences shown with asterisks in Fig.~\ref{aiphe}). Because overshooting increases the main-sequence lifetime, but accelerates the crossing of the subgiant branch, the constraints provided by AI~Phe will remain valuable for the next version of the models in which the mixing length parameter, overshooting, and He diffusion will be tuned together to match all the available observational constraints. 

Two aspects deserve mention in this discussion: {\it (i):} What makes AI~Phe so valuable for stellar evolution is the tiny error of the masses (only 0.4\%); with mass errors of 3\% the test would be far less conclusive. And {\it (ii):} The Yonsei-Yale models yield a mean age for AI~Phe of 5.0~Gyr, while the experimental Victoria models give 4.1~Gyr -- a non-negligible uncertainty in the context of age determinations for nearby solar-type stars. The still-uncertain features of stellar evolution models that can be usefully tested with the best binary data are therefore not just of purely academic interest, but have significant implications for Galactic research as well.

\subsection{Current topics in stellar evolution models}

\paragraph{Convective overshooting.}
A91 discussed the comparison of stellar models with the binary data presented in that paper, with special focus on the then much-debated subject of overshooting from convective cores. A reasonable consensus that the phenomenon is indeed real developed in the following decade, and some formulation of convective overshooting is now incorporated in most models for the evolution of stars above $\sim$1.15 $M_{\odot}$ that have been published since then. 

More recently, attention has turned to the physical description of overshooting and calibration of the parameter(s) describing its extent in stars of different mass and metallicity \citep[see, e.g.,][]{Claret:07}. The mass range 1.1--1.5 $M_{\odot}$ is of special interest here, as the convective core is small and overshooting potentially relatively important. Stars in this mass range are also those that are used as age tracers in the range of interest for Galactic evolution studies, 1--10~Gyr. The effects of slight changes in the amount of overshooting or the detailed elemental composition of the stars are most evident in the detailed morphology of the main-sequence turnoff, which is more clearly revealed by well-populated cluster sequences \citep{davb07} than by the point probes provided by binary systems (but note the discussion of AI~Phe above). Accordingly, recent stellar evolution models have tended to calibrate their overshooting prescription from studies of stellar clusters -- see, e.g., \cite{davb06}. The ongoing programmes to obtain accurate data for eclipsing binaries that are members of well-studied clusters will offer particularly strong constraints on the next generation of models by combining the power of both approaches -- see, e.g., \cite{2008A+A...492..171G}.

Current discussions of stellar models also focus on the applications of asteroseismology, which led to the direct demonstration of the gravitational settling of He in the outer layers of the Sun. One of the reasons for the recent spectacular success of helioseismology is, in fact, that models are also constrained by the accurate values for the solar mass, radius, luminosity and chemical composition. Asteroseismology of other stars is, however, rarely performed on eclipsing binaries, hence on the stars with the most accurate masses and radii. While there are obvious practical reasons for this, the information content of the comparison of asteroseismic models with data for single stars is limited by our uncertain knowledge of their masses and radii. 

\paragraph{Activity in low-mass stars.}
Recent results on G-K-type eclipsing binaries have demonstrated unambiguously that stars with masses just below that of the Sun and in short-period binary orbits exhibit major discrepancies from standard stellar models that provide satisfactory fits to similar stars in long-period orbits, hence with slower rotation \citep{popper, JVC, v1061cyg}. The effect is even more obvious for M stars \citep{yygem, cucnc, guboo}. In short, these stars are up to 10\% larger than their slowly-rotating counterparts -- a huge effect when compared to observational errors of $\sim$ 1\% -- and up to $\sim$400~K cooler. The two effects combine to yield the same luminosity as normal stars, indicating that this is a surface phenomenon.  Early hints of similar discrepancies go back at least 30 years \citep{hoxie73, lacy77}, but the recent accurate results have removed any remaining observational ambiguity.

A shown convincingly by \cite{v1061cyg}, \cite{L-M}, \cite{morales}, and others, these effects are caused by significant surface activity (spots) on the faster-rotating stars. This is evident not only in the light curves, but also in the emission cores of the Ca~II H and K lines and sometimes in X-rays as well. The accepted cause is strong surface magnetic fields, which inhibit efficient convection; and indeed, models with artificially low values of the mixing-length parameter in the outer convective zone (i.e., less efficient convection) fit the observations considerably better \citep{v1061cyg, chabrier, v636cen}. Metallicity has been discussed as an additional cause \citep{berger, L-M}, but would not be expected to affect fast and slow rotators differently.

It would appear that progress in observational accuracy has revealed a class of mildly-active binary stars, intermediate between ordinary inactive stars and the more extreme class of RS~CVn binaries, but with properties clearly different from normal single stars. It remains to be seen whether a continuum of properties exists between these classes of stars, but new models are clearly needed which take these phenomena into account in a physically realistic way. Some headway on this front has already been made \citep[see, e.g.,][]{dantona, mullan, chabrier}.

\section{Tidal evolution and apsidal motion}\label{apsidal}

Well-detached binaries with accurate absolute dimensions provide excellent data with which to study the dynamical effects of tidal friction as well as to explore the internal stellar structure. Tidal evolution is observed by measuring the degree of circularisation of the orbit and the level of synchronisation of the rotational velocities, being a very active field with discussions on alternative theories for the physical description of tidal friction \citep[see, e.g.,][and references therein]{mazeh}. 

Internal structure constants $\log k_2$ are indicative of the degree of central density concentration of the component stars and can be observed in eccentric systems by measuring the apsidal motion period \citep[e.g.,][]{2007IAUS..240..290G}. In Table~\ref{tableaps} we list all systems from Table~\ref{tableMR} with eccentric orbits, as well as those with measured apsidal rates d$\omega$/dt. References are given for the apsidal motion determinations. In three cases (EW~Ori, V459~Cas, and MY~Cyg), the original values were corrected to an adopted eccentricity consistent with the photometric and spectroscopic studies. Here we do not attempt to perform a detailed analysis of the individual systems in this table, but rather to provide a high-quality database satisfying the adopted selection criteria, allowing such studies, including the confrontation with stellar evolution models.

\subsection{Tidal circularisation and synchronisation}\label{tidalevol}

\begin{figure}[ht]
\center \includegraphics[width=\textwidth]{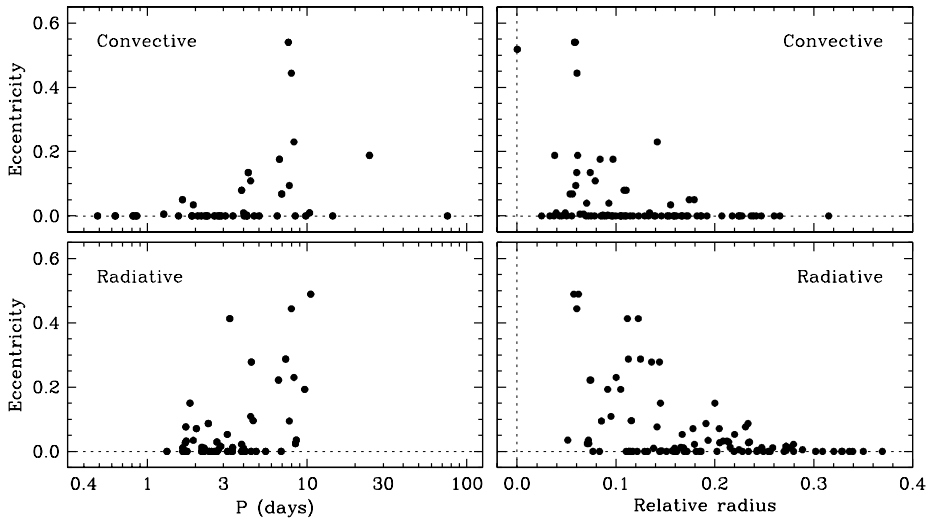}
\caption{Eccentricity as a function of period (left panels) and relative radius (right) for all systems in Table~\ref{tableMR} (except $\alpha$~Cen and OGLE~051019). Individual stars with convective and radiative envelopes are shown separately, with $T_{\rm eff} = 7000$~K as the dividing line.}
\label{tidal}
\end{figure}

Our sample of detached binaries contains both circular and eccentric orbits; in fact, 44 of the 95 systems are eccentric. The left-hand panels of Fig.~\ref{tidal} show the distribution of orbital eccentricity as a function of orbital period, separately for stars with radiative and convective envelopes, adopting $T_{\rm eff} = 7000$~K as the limit between the two groups. For clarity, the two longest-period systems, $\alpha$~Cen and OGLE~051019, are not shown.

Observational biases limit the sample to orbital periods mostly below 10 days. The special case of TZ~For, with two evolved stars in a circular orbit of period 75.7 days, was explained in detail by \cite{1995A+A...296..180C}, who integrated the circularisation time scales along the evolution of the component stars. The only other longer-period system in this diagram, AI~Phe (24.6 days), shows an eccentric orbit, but with synchronised rotational velocities.

As expected, short-period systems present circular orbits while the longer-period binaries show a wide range of eccentricities: No eccentric orbit is found for periods below 1.5 days. It is also clear that systems with stars having convective envelopes circularise more easily, and up to longer periods, than those with radiative envelopes. Diagrams such as Fig.~\ref{tidal} are often used because periods and eccentricities are easily obtained, even for non-eclipsing systems. However, our data also allow us to plot the orbital eccentricity as a function of relative radius (i.e., the radius of the star in units of the orbital semi-axis major). We do so in the right-hand panels of Fig.~\ref{tidal}, which is more interesting from a physical point of view, given the dependence of tidal circularisation time scales on high powers of the relative radii. 

Our sample clearly shows a decreasing dispersion in eccentricity with increasing relative radius, all orbits being circular for relative radii above $\sim$0.25. The long-period system $\alpha$~Cen fits naturally into Fig.~\ref{tidal} as an eccentric system with near-zero relative radii. Again, convective envelopes achieve circularisation for smaller relative radii than radiative ones: Highly eccentric orbits are observed only for quite small relative radii in stars with convective envelopes, while circular orbits are already rare among radiative stars below relative radius 0.1.

\begin{figure}[ht]
\center \includegraphics[width=0.7\textwidth]{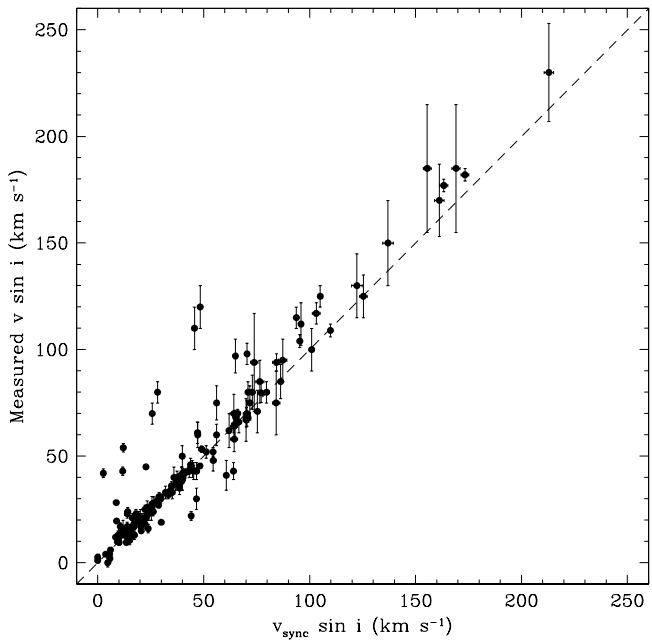}
\caption{Measured $v \sin i$ as a function of the value expected for orbital synchronisation (equal rotational and orbital periods). The one-to-one relation is shown as a dashed line, for reference.}
\label{vsini}
\end{figure}

The other stellar parameter of importance in understanding tidal evolution in binary systems is the level of synchronisation of the component stars. Fig.~\ref{vsini} shows the relation between the observed values of $v \sin i$ and those expected from synchronisation with the orbital period. Some deviating cases are found, taking observational uncertainties into account, mainly in the sense of the observed rotations being faster than synchronous, but some cases of sub-synchronous rotation are also seen. Only the high quality of our data allows to identify these non-synchronous cases with confidence. 

In eccentric systems, tidal forces vary over the orbital cycle, being strongest at periastron. One thus expects the stars to rotate at a rate intermediate between the orbital angular velocity at periastron and that expected for similar single stars. The rotation period of single stars is generally shorter than the typical orbital period of the systems in Table~\ref{tableMR} for early-type stars with radiative envelopes, longer for late-type stars with convective envelopes. The speed with which the stars are spun up or down to their final rotational velocity will, of course, depend on the strength of the tidal forces in each system, i.e., primarily on the relative radii of the stars.

\begin{figure}[t]
\center \includegraphics[width=0.8\textwidth]{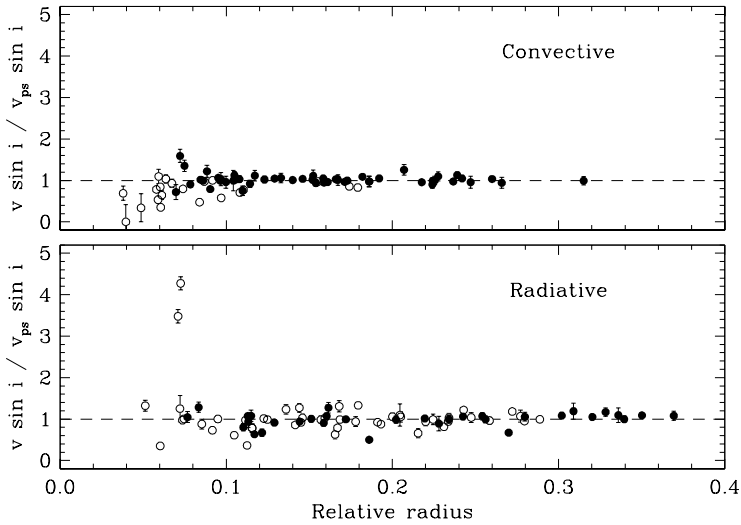}
\caption{Ratio between measured and projected (pseudo-)synchronous rotational velocities for stars with convective (top) and radiative envelopes (bottom). Stars in circular and eccentric systems are shown by filled and open circles, respectively. }
\label{vps}
\end{figure}

Calculations of the average effect of the tidal forces over an eccentric orbit lead to a prediction of the final net rotation of the components -- the concept of pseudo-synchronisation as defined by \cite{1981A+A....99..126H}. Taking this as the best average prediction for the observed rotation rates, Fig.~\ref{vps} shows the level of pseudo-synchronisation achieved by the stars as a function of relative radius (pre-main-sequence stars excluded). For the sake of clarity, stars with convective and radiative envelopes are shown in separate panels, and circular and eccentric orbits by different symbols. 

As seen, pseudo-synchronisation is in fact an excellent approximation for the great majority of the stars; most of the non-synchronous cases are found below relative radius 0.1, as expected from tidal evolution theory. Interesting exceptions are V459~Cas (no.\ 44 in Table~\ref{tableMR}) in the bottom panel of Fig.~\ref{vps} at relative radius $\sim$0.07, with stars rotating much faster than predicted, and the slightly eccentric RW~Lac (no.\ 90) in the top panel of Fig.~\ref{vps} at relative radius 0.04--0.05, with highly sub-synchronous rotation. These exceptions are consistent with the expected levels of synchronisation for such small relative radii, given that the overall tendency in binaries is to slow down radiative stars and spin up convective ones relative to their single counterparts. 

The good quality of the data, as reflected in the near-invisible error bars, reveals a smaller dispersion among stars with convective rather than radiative envelopes due to the more efficient circularisation mechanism. The data also reveal a number of sub-synchronous radiative stars with relative radii above 0.1 that cannot yet be explained (e.g., V451~Oph or V1031~Ori). Detailed stellar models with integrated tidal evolution calculations will be needed to address this issue.

Using different symbols for stars in eccentric and circular orbits in Fig.~\ref{vps} also reveals any effects of overcorrection for eccentricity in convective stars when adopting pseudo-synchronisation. EY~Cep \citep{eycep} is an interesting case of a highly eccentric young binary where theory predicts a non-circular orbit as observed, but not that the two stars should rotate at the orbital rate. Presumably, synchronisation was achieved by deep convective envelopes, now or during the pre-main sequence phase, or the stars managed to slow down independently. In other cases, the two stars in the same system show clearly different behaviours. For example, the primary of V364~Lac rotates faster than expected, while the secondary rotates sub-synchronously; the opposite is the case in V396~Cas. 

Overall, the normal pattern is that the larger primary stars are synchronised while the smaller secondaries are still on their way to the final state, either from faster rotation in radiative stars ($\zeta$~Phe) or from slower speeds in stars with convective envelopes (BW~Aqr). Nevertheless, cases exist that require special attention, including new observations and additional tidal modelling (e.g. V539~Ara or CV~Vel). For clarity, the exceptionally fast-rotating secondary component of TZ~For (spinning more than 15 times above the synchronous rate) has been excluded from Fig.~\ref{vps}; the tidal history and special evolutionary configuration of this system have been studied in detail by \cite{1995A+A...296..180C}. Recent advances in tidal theory include the work of \cite{Kumar:96}, \cite{Witte:99a, Witte:99b, Witte:02}, and \cite{Willems:03}. An excellent summary of the topic and its applications can be found in the proceedings of the 3rd Granada Workshop on Stellar Structure \citep{Claret:05}.

\subsection{Apsidal motion}

Important additional information about stellar structure is available if the rate of apsidal motion in an eccentric binary system can also be measured. This is the case for 29 of the 44 eccentric systems in our sample, although for two of them the apsidal motion has not been measured with enough precision to allow for a significant comparison with theory. One of these systems is BP~Vul \citep{2003AJ....126.1905L}; the other is the extremely interesting case of CM~Dra, with the lowest stellar masses of the sample \citep{cmdra}. 

The much-discussed system DI~Her requires special mention \citep[see][]{Claret:98}. DI~Her was excluded from our overall study because of the recent discovery, based on the Rossiter effect, that the spin axes of the stars are almost perpendicular rather than parallel to that of the orbit \citep{albrecht}. This configuration is, in fact, not unlikely in such a young binary with small relative radii, provided that the stars were initially formed with misaligned spin axes. In summary, when the observed misalignment, the Shakura effect \citep{1988ApJ...335..962C}, and the general relativistic contribution are accounted for in the tidal and rotational terms of the predicted apsidal motion, excellent agreement is obtained with the observed apsidal motion rate. 

The observed apsidal motion in binary stars has two contributions due to the non-Keplerian dynamical behaviour of the component stars. The classical term is caused by the stellar distortions produced by rotation and tides, while the non-Newtonian term corresponds to the predictions of General Relativity. For a recent comparison between observed and predicted apsidal motion rates, including for the first time the effects of dynamical tides, see \cite{Claret:02}.

\begin{figure}[ht]
\center \includegraphics[width=0.65\textwidth]{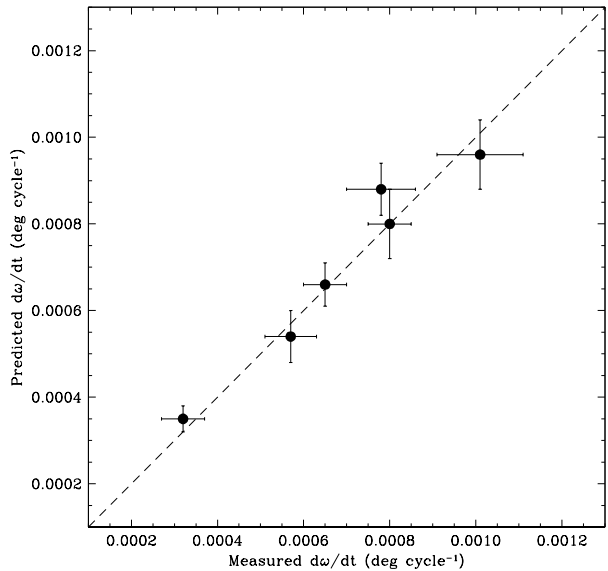}
\caption{Observed vs.\ predicted apsidal motion rates for systems with a general relativistic contribution of at least 40\% of the total. The systems are, from left to right, EW~Ori, V459~Cas, GG~Ori, V1143~Cyg, V636~Cen, and EK~Cep.}
\label{apsidal1}
\end{figure}

\paragraph{The general relativistic term.}
The relativistic term of the apsidal motion was examined, e.g., by \cite{Claret:97}, and reviewed more recently by \cite{2007IAUS..240..290G}, who analysed 16 systems. He found good agreement between predicted and observed rates for 12 of them, but the sample was not of the same quality as that presented here. For our comparison between predicted and observed apsidal motions, we have considered all the systems in Table~\ref{tableMR} with eccentric orbits and well-determined apsidal motions, and with a predicted relativistic contribution of at least 40\% of the observed total rate. This allows us to minimize the influence of possible errors in the models used to compute the tidal contribution. 

Only six systems fulfil these conditions, excluding the special case of DI~Her, discussed above. For these systems, the expected apsidal motion rates were computed including the general relativistic contribution, using theoretical models of internal structure as described below, and the observed rotational velocities (see Table~\ref{tableMRsup}). The predicted and observed apsidal motion rates, in degrees per cycle, are compared in Fig.~\ref{apsidal1}. The good agreement seen there, together with the resolution of the `DI~Her enigma', seems to indicate that apsidal motion as an argument in favour of alternative theories of gravitation is a closed case.

\paragraph{The tidal terms.}

Applying the standard correction for the general relativistic contribution to the observed apsidal motion for the rest of the systems in Table~\ref{tableaps},  we can compute the average internal structure constant, $\log k_2$. In order to ensure that the resulting values are of good accuracy, we consider only systems where the tidal and rotational effects contribute at least 40\% of the observed total apsidal motion. Moreover, pinpointing individual main-sequence systems between the ZAMS and the TAMS (terminal-age main sequence) requires a precision of at least 0.1 in $\log k_2$. Only 18 binaries satisfy these quality criteria, and we list the derived values of $\log k_2$ for these stars in Table~\ref{tableaps}.

\begin{figure}[t]
\center \includegraphics[width=\textwidth]{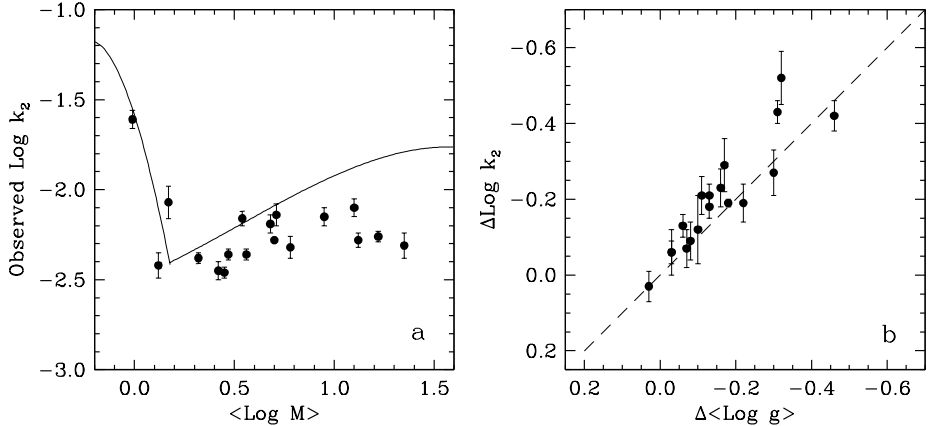}
\caption{{\it (a)} Measured internal structure constant as a function of the weighted mean logarithmic mass in each system. The ZAMS relation from the \cite{1995A+AS..109..441C} models for solar metallicity is shown. {\it (b)} Deviations of the measured internal structure constant from the ZAMS relation in {\it (a)} as a function of the weighted mean surface gravity in each system. The dashed line is for reference; no physical relation is suggested.}
\label{apsidal2}
\end{figure}

Fig.~\ref{apsidal2}a shows $\log k_2$ for these systems as a function of the mean mass of the stars (using the same weighting procedure as implicit in the observed structure constant). Theoretical values from ZAMS models for the solar chemical composition from \cite{1995A+AS..109..441C} are also shown. It is clear that the precision of the data allows us to follow the stars as they evolve beyond the ZAMS, towards smaller values of $\log k_2$ (greater central concentration); note that the lower-mass stars are generally less evolved. No correction for variation in metal content has been made in this general plot. 

That the downward shift of the points in Fig.~\ref{apsidal2}a is primarily due to evolution is clearly seen in Fig.~\ref{apsidal2}b, which shows the difference between the observed and theoretical (ZAMS) $\log k_2$ values as a function of the difference in the observed and ZAMS values of the mean surface gravity, $\Delta\log g$. The nearly linear correlation and increasing dispersion in $\Delta\log k_2$ with increasing $\Delta\log g$ were suggested already by the evolutionary models of \cite{1995A+AS..109..441C} (his Fig.~7). Corrections for mild degrees of evolution, derived from this relation, were applied to the computed tidal contributions when assessing the general relativistic terms above.

Given that differences in metal content or rotation of the stars were not taken into account, the observed tight correlation is encouraging. Computing specific models for the observed mass, chemical composition and degree of evolution of each of the component stars would no doubt provide useful constraints on the adopted input physics when compared with these observations. The next few years should see a significant increase in the number of systems with reliably determined internal structure constants.  

\section{Calibration of single-star properties}\label{singlestars}

With our new sample of nearly 200 accurate masses and radii, the question arises if a calibration can be devised to estimate precise values of $M$ and $R$ for single stars from observable indicators of the basic parameters mass, composition, and age. A91 speculated about the way such a calibration could be devised, but did not proceed to action on the basis of the data available then. Our new, larger sample of binary parameters -- including [Fe/H] for many systems -- allowed us to make a new attempt, described in the following.

\begin{figure}[ht]
\center \includegraphics[width=0.8\textwidth]{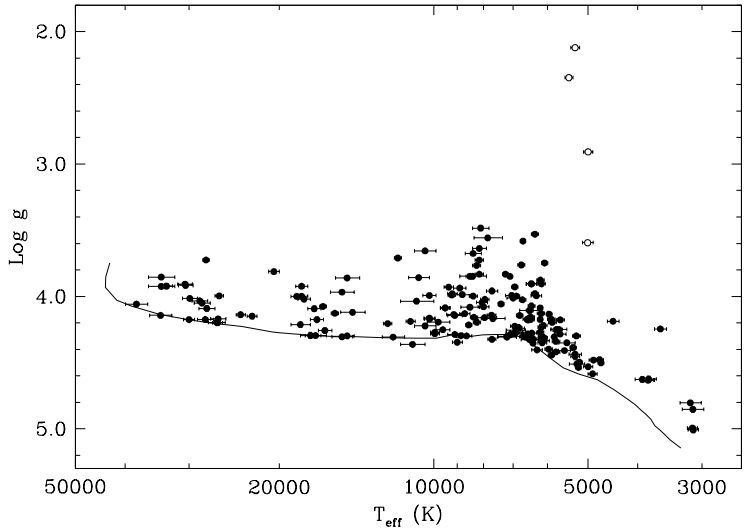}
\caption{Log $g$ vs.\ $T_{\rm eff}$ for the stars in Table~\ref{tableMR}, together with a 1-Myr isochrone for $Z = 0.019$ from \cite{girardi}. Error bars are shown. Open circles: Stars classified as giants.}
\label{logglogT}
\end{figure}

\paragraph{Calibrations of M and R.}
Fig.~\ref{logglogT} highlights the degree of evolution of each star in the sample away from the ZAMS, which is roughly horizontal in this diagram, as a function of temperature ($\simeq$ spectral type). Evolution without mass loss is generally upwards and to the right in this figure, which suggests that suitable indicators of the degree of evolution of a star are $T_{\rm eff}$ and $\log g$, which are observable by both spectroscopic and photometric techniques, together with the metal abundance [Fe/H]. 

\begin{figure}[ht]
\center \includegraphics[width=0.8\textwidth]{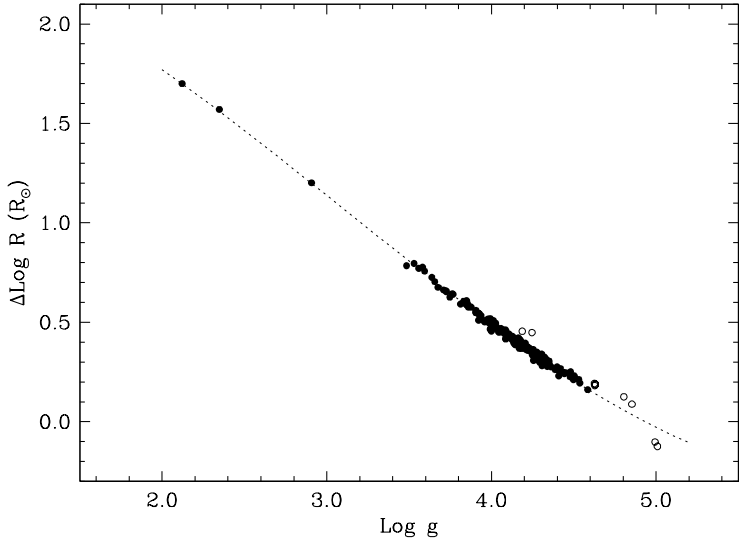}
\caption{Deviation of the observed radii from a polynomial ZAMS relation in $T_{\rm eff}$ and [Fe/H], $\Delta \log R$ vs.\ $\log g$. Open symbols denote stars below 0.6~$M_{\odot}$ and pre-main-sequence stars. The dotted line represents the remaining fitted dependence on $\log g$.}
\label{deltaR}
\end{figure}

Accordingly, we have attempted to model first the radii of the stars in Fig.~\ref{logRlogT} by fitting a ZAMS relation, then fitting the deviations from that relation as functions of $\log g$, with small additional terms in [Fe/H]. That strategy was quite successful, as illustrated in Fig.~\ref{deltaR}, which shows the correlation between $\log g$ and the deviation of the observed radii from a global polynomial fit including terms in $T_{\rm eff}$ and [Fe/H] only (i.e. not just a ZAMS fit to Fig.~\ref{logRlogT}). Note that stars below 0.6~$M_{\odot}$ and pre-main-sequence stars (open circles) do not fit this relation and are excluded from the following discussion.

We therefore proceeded to perform a full fit to $M$ and $R$, expressed as the simplest possible polynomials in $T_{\rm eff}$, $\log g$ and [Fe/H]. The resulting equations are given below and the coefficients listed in Table~\ref{tableCoef}, with one extra guard digit. 
\begin{eqnarray*}
\log M &=& a_1 + a_2 X + a_3 X^2 + a_4 X^3 + a_5 (\log g)^2 + a_6 (\log g)^3 + a_7 {\rm [Fe/H]} \\
\log R &=& b_1 + b_2 X + b_3 X^2 + b_4 X^3 + b_5 (\log g)^2 + b_6 (\log g)^3 + b_7 {\rm [Fe/H]}~,
\end{eqnarray*}
where $X = \log T_{\rm eff} - 4.1$. The scatter from these calibrations is $\sigma_{\log M} = 0.027$ and $\sigma_{\log R} = 0.014$ (6.4\% and 3.2\%, respectively) for main-sequence and evolved stars above 0.6~$M_{\odot}$. The larger error in the mass, which varies over a smaller range than the radius, suggests that mass may depend on the input parameters in a more complex way than that described by these equations (see also below).

\begin{table}
\begin{center}
\caption{Coefficients for the calibration equations above.\label{tableCoef}}
\begin{tabular}{cc@{}c@{}}
\noalign{\smallskip}\hline\noalign{\smallskip}
$i$ & $a_i$ & $b_i$ \\
\noalign{\smallskip}\hline\noalign{\smallskip}
1 & 1.5689~$\pm$~0.058\phn        & 2.4427~$\pm$~0.038\phn \\
2 & 1.3787~$\pm$~0.029\phn        & 0.6679~$\pm$~0.016\phn \\
3 & 0.4243~$\pm$~0.029\phn        & 0.1771~$\pm$~0.027\phn \\
4 & 1.139~$\pm$~0.24\phn          & 0.705~$\pm$~0.13\phn \\
5 & $-$0.1425~$\pm$~0.011\phn\phs & $-$0.21415~$\pm$~0.0075\phn\phs \\
6 & 0.01969~$\pm$~0.0019\phn      & 0.02306~$\pm$~0.0013\phn \\
7 & 0.1010~$\pm$~0.014\phn      & 0.04173~$\pm$~0.0082\phn \\
\noalign{\smallskip}\hline\noalign{\smallskip}
\end{tabular}
\end{center}
\end{table}

\begin{figure}[ht]
\center \includegraphics[width=\textwidth]{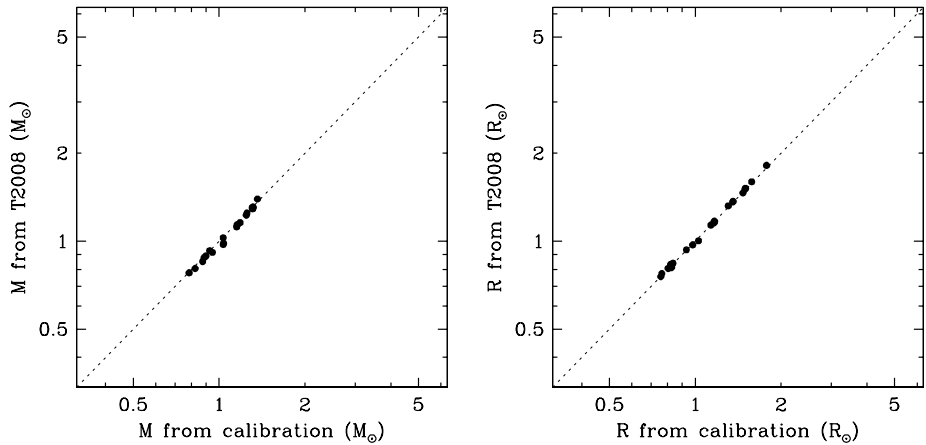}
\caption{Masses and radii as derived in this paper vs.\ those derived for host stars of transiting planets by \cite{GT08}.}
\label{CalibTest1}
\end{figure}

\paragraph{Testing and applying the calibrations.}
In principle, the above calibrations allow one to infer the mass and radius of a single star to a few per cent from an observed set of accurate values of $T_{\rm eff}$, $\log g$ and [Fe/H]. For example, when applied to the Sun itself, the results are $M = 1.051~M_{\odot}$ and $R = 1.018~R_{\odot}$, well within the scatter of the calibrations. These relations are particularly interesting for exoplanet host stars, where the properties of the planet are normally obtained relative to those of the star. We have therefore checked the results of our purely empirical calibrations with the set of results obtained for the host stars of transiting planets by \cite{GT08}. For such stars, additional information about the radius and $\log g$ is available from the transit light curves, and masses are then inferred from stellar evolution models. The results of the comparison are shown in Fig.~\ref{CalibTest1}, and the agreement is very satisfactory.

\begin{figure}[ht]
\center \includegraphics[width=\textwidth]{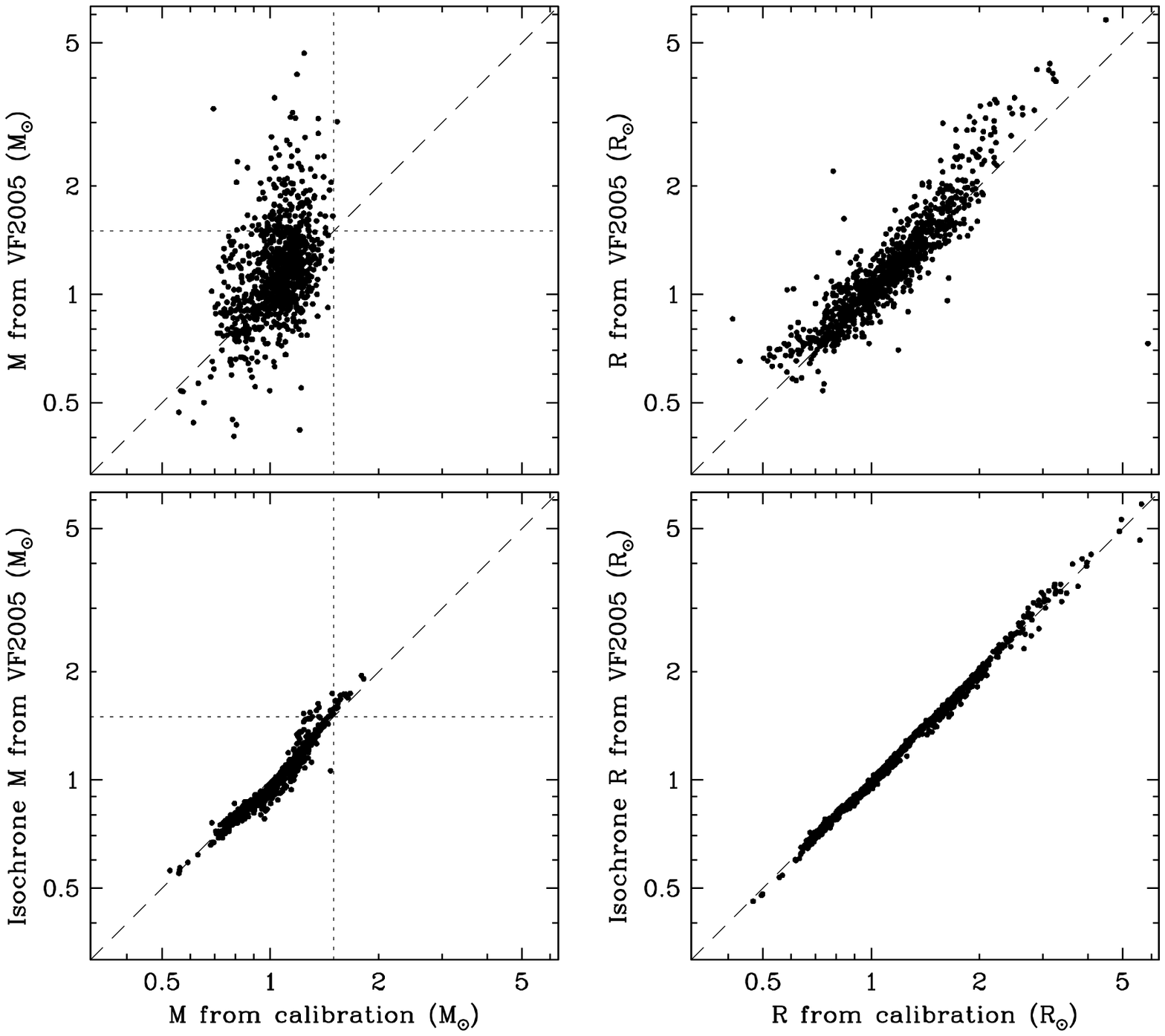}
\caption{Masses and radii as derived in this paper vs.\ those derived for 1,040 F, G, and K stars by \cite{VF05}. {\it Top panels:} Mass and radius as derived directly from the parallax-based luminosity, $T_{\rm eff}$, and $\log g$.  {\it Bottom panels:} Mass and radius as derived from theoretical isochrones and luminosity, $T_{\rm eff}$, and [Fe/H] as input parameters. Dotted lines in the left-hand panels indicate the approximate mass of an F0 star, the largest to be expected in this sample.}
\label{CalibTest2}
\end{figure}

To further explore the potential of our new calibrations, we have derived masses and radii for the 1,040 nearby F, G, and K stars studied by \cite{VF05} from their spectroscopic determinations of $T_{\rm eff}$, $\log g$ and [Fe/H]. Their masses and radii are derived in two ways: either directly, i.e., the radius from luminosity based on the Hipparcos parallax \citep{Perryman:97} combined with $T_{\rm eff}$, then the mass from $\log g$, or alternatively, $M$ and $R$ are derived from theoretical isochrones with the luminosity and spectroscopic $T_{\rm eff}$ and [Fe/H] values as input parameters. Their results and our values are compared in Fig.~\ref{CalibTest2}. 

Two features are prominent: First, the masses and radii derived from isochrones by \cite{VF05}, and preferred by them, are indeed much more reliable than those derived directly from parallax, $T_{\rm eff}$ and $\log g$; in particular, implausibly large masses are found for the evolved stars in the latter case. Second, a small, but significant deviation ($\sim$5\%) is seen for masses near $1~M_{\odot}$, in the sense that our calibrations give slightly larger masses than the isochrones. Because the same effect is found for the Sun, as noted above, the isochrones are probably not the cause of this difference. 

We have attempted to refine our mass calibration with higher-order terms, but without success. The number of systems with good [Fe/H] determinations is still too small to support a more sophisticated approach, so we prefer to retain the simple equations above, noting that the accuracy achieved is still very good and the equations far simpler to use than interpolating in isochrone tables.

\section{Systems with accurate interferometric masses}\label{Mdata}

Progress in long-baseline optical interferometry of close visual binaries has resulted in an increasing number of systems with accurate interferometric and spectroscopic orbits. Table~\ref{tableMonly} lists the 23 systems in the literature in which the individual stellar {\it masses} have been determined to better than the limit of 3\% for stars included in this review (Tables~\ref{tableMR}--\ref{tableMRsup}). Again, we have examined the original material on which these masses are based and satisfied ourselves in each case -- if necessary by independent orbital computations -- that the error estimates we list are indeed reliable. The orbital parallaxes also yield accurate distances, which we list in Table~\ref{tableMonly} as well, along with other information such as metallicity, when available.

\begin{figure}[th]
\center \includegraphics[width=0.8\textwidth]{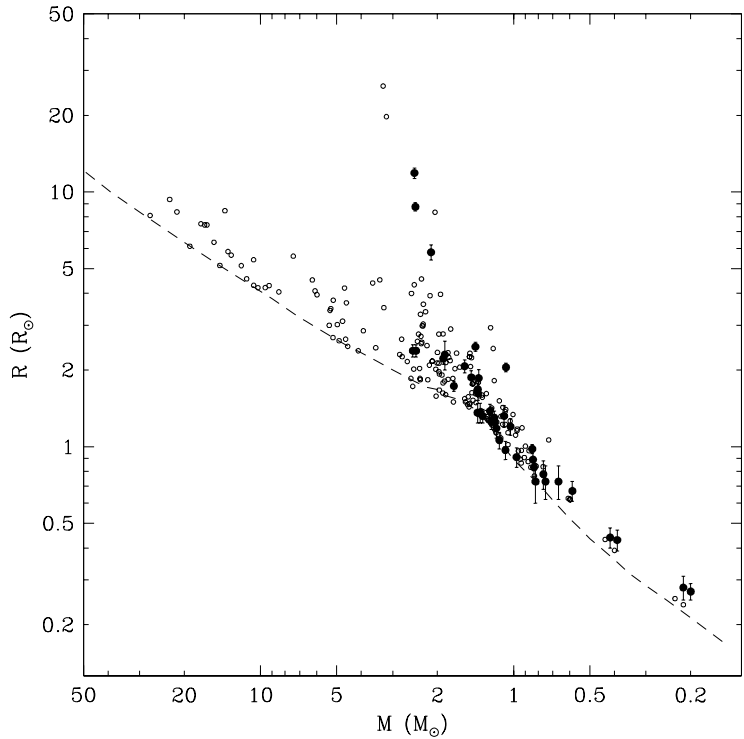}
\caption{Mass-radius diagram for the interferometric binaries in Table~\ref{tableMonly} (filled circles), together with those from Table~\ref{tableMR} (open circles). The ZAMS relation is the same as shown in Fig.~\ref{logRlogM}. See text for the determination of $M$ and $R$.}
\label{logRlogMinterf}
\end{figure}

However, only for Capella ($\alpha$~Aur) have the stellar radii been determined from directly measured angular diameters and parallaxes, and even then not yet with the accuracy we require. For the other systems in Table~\ref{tableMonly}, radius estimates are based on observed luminosities (from apparent magnitudes and distances) and temperatures. Such estimates are indirect and scale-dependent, and their relative errors -- twice those in the temperature -- exceed our limit of 3\%. Indeed, in several cases the available information on the {\it individual} temperatures and luminosities is too fragmentary for any meaningful estimate of individual radii. 

Fig.~\ref{logRlogMinterf} shows the mass-radius diagram for the interferometric binaries in Table~\ref{tableMonly}, with the stars in Table~\ref{tableMR} shown as well for comparison. Two salient features appear: The much larger error bars in $R$ than seen in Fig.~\ref{logRlogM}, and the addition of significant numbers of low-mass and evolved stars. The low accuracy of the radii prevents us from including them in the types of discussion contained in the previous sections, but we list them here as a stimulus to observers to complement the excellent mass determinations with the missing essential data of matching quality.

Finally, we note that a fairly large number of single stars exist for which accurate absolute radii have been determined from angular diameters and parallaxes. However, as mass values of matching credentials cannot be determined for these stars, we have decided to not discuss them in the context of this review.

\section{Directions for future work}\label{future}

Based on our assessment of the current status of our knowledge of accurate stellar masses and radii, we point out in the following a number of directions in which further work appears especially promising.

\paragraph{Coverage of the stellar parameter space.} 
Relative to the sample of A91, the number of massive stars ($M > 10~M_{\odot}$) has increased from 6 to 17, but only one star more massive than the previous record holder has been added in these 18 years. Similarly, the number of stars less massive than the Sun has increased from 5 to 25, but only four are less massive than YY~Gem. And only one pre-main-sequence system and one giant system (in the LMC) have been added since the earlier review. Additional studies of these types of star do exist, but refinement of the stellar parameters to the level adopted here is necessary for these systems to become truly useful. We note that optical interferometry will be valuable in determining masses for low-mass and giant stars, but radius determination of matching accuracy remains an issue.

\paragraph{Effective temperatures.} 
$T_{\rm eff}$ is a key parameter in all discussions of stellar and Galactic evolution, directly affecting the location of a star in the HR diagram and the use of a star to determine distances to other galaxies or age scales of galactic populations. Given the current disagreement between several spectroscopic and photometric temperature scales (see, e.g., \citealt{GCS2} for a detailed discussion), improvement of the $T_{\rm eff}$ scale via additional accurate angular diameter and flux measurements is the most urgent priority. In the process, the interstellar reddening must be carefully determined for both programme stars and calibrators.

\paragraph{Metallicity.} 
As seen in Table~\ref{tableMRsup}, measurements of [Fe/H] still exist for only a minority of the stars discussed here; more detailed abundances for even fewer. For all serious determinations of stellar ages -- and indeed for most astrophysical discussions of these stars -- the chemical composition is a key parameter. While acknowledging that the analysis of double-lined spectra is more challenging than for single stars, we point out that modern tomographic or disentangling techniques are now available to facilitate the task. Chemical composition data are particularly urgently needed for the low-mass stars, for which current models are the most uncertain.

\paragraph{Rotation.} 
Accurate values of $v\sin i$ are needed in order to verify to what extent real binary components rotate as predicted by stellar and tidal evolution theory. Some outliers are explained by the stars being too young and/or too widely separated for tidal synchronisation to have been fully effective, but in other cases other effects may play a role, as suggested by Fig.~\ref{vps}. Clarification of such cases may require that $v\sin i$ be redetermined on a homogeneous basis from modern high-quality spectra.

\paragraph{Stellar models.} 
Better stellar evolution models will be needed to take full advantage of the data presented here, especially for low-mass and active stars below 1 $M_{\odot}$. For the latter, models must address the influence of strong, rotation-generated magnetic fields and large-scale surface inhomogeneities (spots) that affect the radius and luminosity of the star significantly. For stars in the 1.1--1.5 $M_{\odot}$ range, precise and tested prescriptions for the combined effects of core overshooting and He diffusion are needed to further consolidate the determination of stellar ages throughout the lifetime of the Galactic disk. 

\section{Conclusions and outlook}\label{TheEnd}

The aim of this paper has been to summarise the status of accurate determinations of stellar masses and radii, and of the quantities that are needed for the astrophysical discussion of these data. The use of the data in checking models of stellar evolution and tidal interaction in binaries was discussed in some detail in Sect.~\ref{models}--\ref{tidalevol}, and calibrations relating the observed $T_{\rm eff}$, $\log g$ and [Fe/H] of a star directly with its mass and radius were derived in Sect.~\ref{singlestars}. Sect.~\ref{Mdata} lists a number of interferometric binaries with accurate masses, but still with insufficiently precise radii and other quantities. Finally, Sect.~\ref{future} points out a number of directions for future work in the field. Rather than recapitulate these results, we wish to end the paper by taking a look at the future.

A new era of large-scale photometric surveys from space of unprecedented accuracy has dawned on us with the launch and successful operation of the CoRoT and Kepler space observatories. The boom in searches for transiting extrasolar planets -- which had not even begun at the time of the \cite{A91} review -- will ensure that such surveys will continue in the foreseeable future. The prospect is an influx of tens of thousands of well-covered light curves of binary and other variable stars with photometric errors 2--3 orders of magnitude below those discussed in this paper. What will this mean?

A first consequence will be that a thorough revision of the existing light curve analysis codes will be needed: None of them is developed to model the effects of stellar deformations, disk intensity distributions (including spots), and reflection and scattering of light in the system to the level of a few parts per million. A second consequence will be increased demands on the spectroscopic follow-up, not only for more accurate mass determinations (which will also have to consider proximity effects that are now negligible), but also for the more accurate and detailed abundance determinations which will be needed to interpret the more accurate masses and radii.

In turn, this will send the ball back in the court of the stellar evolution modellers, as the observations will allow us to test the importance of a next level of effects that are ignored in the current generation of models, such as magnetic fields, non-rigid rotation, chemical fractionation, surface inhomogeneities, etc. Only when models and a complete set of observations progress hand in hand will the full advance in astrophysical understanding be achieved.

How will this influence the field of asteroseismology, which feeds from the same space data as the extrasolar planet searches? Already a few cases are known \citep[e.g., $\beta$~Hyi and $\beta$~Vir;][]{north07, north09} where an accurate parallax and angular diameter led to an accurate absolute radius for the star which, via a seismological determination of the mean stellar density, led to a precise value for the mass. We did not include these stars in Table~\ref{tableMRsup}, because the mass rests more heavily on theoretical assumptions than we have preferred to do, but such cases can be expected to multiply. The challenge for the future will be to combine the accurate mass and radius determinations from binary systems with the results of asteroseismological analyses, which are best performed on single stars. We expect this to take less than another 18 years.

\begin{acknowledgements}

We thank our main collaborators and friends during many years of research on binary stars, Jens Viggo Clausen and Birgitta Nordstr\"om as well as Andy Boden, Claud Lacy, David Latham, the late Daniel Popper, Ignasi Ribas, Robert Stefanik, and Luiz Paulo Vaz. We also thank Don VandenBerg and Antonio Claret for inspiring collaborations during many years on the theoretical aspects of stellar evolution as illuminated by accurate binary data.
GT acknowledges partial support from NSF grant AST-0708229. JA thanks the Danish Natural Science Research Council, the Carlsberg Foundation, and the Smithsonian Institution for partial financial support for this research, and David Latham for hospitality at the CfA while this paper was being prepared.
This research has made extensive use of the SIMBAD database and the VizieR catalogue access tool, both operated at CDS, Strasbourg, France, of NASA's Astrophysics Data System Abstract Service, and of data products from the Two Micron All Sky Survey (2MASS), which is a joint project of the University of Massachusetts and the Infrared Processing and Analysis Center/California Institute of Technology, funded by NASA and the NSF.

\end{acknowledgements}


{\small
\begin{center}
\begin{landscape}
\begin{longtable}{r@{\thinspace\thinspace\thinspace}l@{~}c@{~~}c@{~~}l@{~}c@{~~~}c@{~~~}c@{~~~}c@{~~~}c@{~~~}cc}

\caption{(continued)} \\
\hline\noalign{\smallskip}
   &     & $P$ (d) &  &  & Mass & Radius
         & $T_{\rm eff}$ & $\log g$ 
         & $\log L$ & $M_V$ \\
& System & $V_{\rm max}$ & Star & SpT & ($M_{\odot}$) & ($R_{\odot}$)
         & (K) & (cgs) & ($L_{\odot}$)
         & (mag)  \\
\noalign{\smallskip}\hline\noalign{\smallskip}
\endhead

\noalign{\smallskip}\hline
\endfoot

\noalign{\smallskip}\hline\noalign{\smallskip}
\multicolumn{12}{l}{Note: References to the original sources of these data are given in Table~\ref{tableMRsup}.}
\endlastfoot

\caption{Observed parameters and derived quantities for 95 binary systems, with mean errors.\label{tableMR}} \\
\hline\noalign{\smallskip}
   &     & $P$ (d) &  &  & Mass & Radius
         & $T_{\rm eff}$ & $\log g$ 
         & $\log L$ & $M_V$ \\
& System & $V_{\rm max}$ & Star & SpT & ($M_{\odot}$) & ($R_{\odot}$)
         & (K) & (cgs) & ($L_{\odot}$)
         & (mag)  \\
\noalign{\smallskip}\hline\noalign{\smallskip}
\endfirsthead

\noalign{\smallskip}
  1 & V3903 Sgr      &  1.74& A & O7V    &  27.27$\pm$0.55\phn&  8.088$\pm$0.086  &38000$\pm$1900\phn & 4.058$\pm$0.016   & 5.088$\pm$0.087       &   $-$4.36$\pm$0.08\phs    \\
    & HD 165921      &  7.27& B & O9V    &  19.01\pppp0.44\phn&  6.125\pppp0.060  &34100\pppp1700\phn & 4.143\pppp0.013   & 4.658\pppp0.088       &   $-$3.64\pppp0.08\phs    \\
\noalign{\smallskip}  
  2 & EM Car         &  3.41& A & O8V    &  22.83\pppp0.32\phn&   9.35\pppp0.17   &34000\pppp2000\phn & 3.855\pppp0.016   & 5.021\pppp0.104       &   $-$4.55\pppp0.10\phs    \\
    & HD 97484       &  8.38& B & O8V    &  21.38\pppp0.33\phn&   8.35\pppp0.16   &34000\pppp2000\phn & 3.925\pppp0.016   & 4.922\pppp0.104       &   $-$4.30\pppp0.10\phs    \\
\noalign{\smallskip}  
  3 & V1034 Sco      &  2.44& A & O9V    &  17.21\pppp0.46\phn&  7.507\pppp0.081  &33200\pppp900\pht  &3.9229\pppp0.0082  & 4.789\pppp0.048       &   $-$4.04\pppp0.06\phs    \\
    & CD$-$41 11042  &  8.32& B & B1.5V  &   9.59\pppp0.27    &  4.217\pppp0.089  &26330\pppp900\pht  & 4.170\pppp0.013   & 3.885\pppp0.062       &   $-$2.44\pppp0.09\phs    \\
\noalign{\smallskip}  
  4 & V478 Cyg       &  2.88& A & O9.5V  &  16.62\pppp0.33\phn&  7.426\pppp0.072  &30479\pppp1000\phn &3.9173\pppp0.0074  & 4.631\pppp0.058       &   $-$3.91\pppp0.07\phs    \\
    & HD 193611      &  8.63& B & O9.5V  &  16.27\pppp0.33\phn&  7.426\pppp0.072  &30549\pppp1000\phn &3.9080\pppp0.0075  & 4.635\pppp0.057       &   $-$3.91\pppp0.07\phs    \\
\end{longtable}
\end{landscape}
\end{center}
}




{\small
\begin{center}
\begin{landscape}
\begin{longtable}{r@{\thinspace\thinspace\thinspace}l@{~~}cccc@{~}c@{}c l}
\caption{(continued)} \\
\hline\noalign{\smallskip}
 &        & Distance & $E(B-V)$
          & $v_{\rm A} \sin i$ & $v_{\rm B} \sin i$
          &        & $\log$ age &            \\
 & System & (pc) & (mag)
          & (km~s$^{-1}$) & (km~s$^{-1}$)
          & [Fe/H] & (yr) & References \\
\noalign{\smallskip}\hline\noalign{\smallskip}
\endhead

\noalign{\smallskip}\hline\noalign{\smallskip}
\endfoot

\noalign{\smallskip}\hline\noalign{\smallskip}
\multicolumn{9}{l}{Notes: An asterisk after the name indicates one or both components are in the pre-main sequence phase. An asterisk after the metallicity} \\
\multicolumn{9}{l}{indicates a value adopted from the binary's parent population (cluster or galaxy).} \\
\endlastfoot

\caption{Other parameters and references for the 95 binary systems in Table~\ref{tableMR}.\label{tableMRsup}} \\
\hline\noalign{\smallskip}
 &        & Distance & $E(B-V)$
          & $v_{\rm A} \sin i$ & $v_{\rm B} \sin i$
          &        & $\log$ age &            \\
 & System & (pc) & (mag)
          & (km~s$^{-1}$) & (km~s$^{-1}$)
          & [Fe/H] & (yr) & References \\
\noalign{\smallskip}\hline\noalign{\smallskip}
\endfirsthead

  1 & V3903 Sgr      &    1389$\pm$73\pht  &     0.44            &   230$\pm$23\phn  &   170$\pm$17\phn  &    \nodata           & 6.3  &  \cite{1997A+A...327.1094V}  \\
\noalign{\smallskip}  
  2 & EM Car         &    2190\pppp120\phn &    0.60             &   150\pppp20\phn  &   130\pppp15\phn  &    \nodata           & 6.6  &  \cite{1989A+A...213..183A}  \\
\noalign{\smallskip}  
  3 & V1034 Sco      &    1711\pppp89\pht  &   0.459             &    \nodata        &    \nodata        &    \nodata           & 6.5  &  \cite{2005A+A...437..769B}  \\
\noalign{\smallskip}  
  4 & V478 Cyg       &    1354\pppp69\pht  &   0.85              &    \nodata        &    \nodata        &    \nodata           & 6.7  &  \cite{1981AJ.....86..102P}  \\
    &                &                     &                     &                   &                   &                      &      &  \cite{1991AJ....101..600P}  \\
    &                &                     &                     &                   &                   &                      &      &  \cite{2006A+A...456.1077W}  \\
\noalign{\smallskip}  
  5 & AH Cep         &     687\pppp82\phn  &   0.58$\pm$0.08     &   185\pppp30\phn  &   185\pppp30\phn  &    \nodata           & 6.7  &  \cite{1986MNRAS.223..513B}  \\
    &                &                     &                     &                   &                   &                      &      &  \cite{1990A+A...236..409H}  \\
\end{longtable}
\end{landscape}
\end{center}
}



{\small
\begin{center}
\begin{longtable}{r@{\thinspace\thinspace\thinspace}l@{~~~}c@{~~}c@{~~}c@{~~~}c@{}c}

\caption{Systems in Table~\ref{tableMR} with eccentric orbits, and systems with apsidal motion.\label{tableaps}} \\
\hline\noalign{\smallskip}
 &        &     & d$\omega$/dt        &            &      &     \\
 & System & $e$ & (deg cycle$^{-1}$)  & $\log k_2$ & Flag & Ref \\
\noalign{\smallskip}\hline\noalign{\smallskip}
\endfirsthead

\caption{(continued)} \\
\hline\noalign{\smallskip}
 &        &     & d$\omega$/dt        &            &      &     \\
 & System & $e$ & (deg cycle$^{-1}$)  & $\log k_2$ & Flag & Ref \\
\noalign{\smallskip}\hline\noalign{\smallskip}
\endhead

\noalign{\smallskip}\hline\noalign{\smallskip}
\endfoot

\noalign{\smallskip}\hline\noalign{\smallskip}
\multicolumn{7}{l}{Flags: 0 - Apsidal motion not detected; 1 - Rotational axes misaligned; 2 - Apsidal} \\
\multicolumn{7}{l}{motion dominated by General Relativity component; 3 - Significance of apsidal} \\
\multicolumn{7}{l}{motion too low for reliable determination of $\log k_2$.} \\
\multicolumn{7}{l}{References to apsidal motion: 1 - \cite{1989A+A...213..183A}; 2 - \cite{2006A+A...456.1077W};} \\
\multicolumn{7}{l}{3 - \cite{1974A+A....34..317W}; 4 - \cite{1986A+A...159..157G}; 5 - \cite{2005A+A...437..545W};} \\
\multicolumn{7}{l}{6 - \cite{uoph}; 7 - \cite{albrecht}; 8 - \cite{2000A+A...356..134W}; 9 - \cite{1993A+A...277..439A};} \\
\multicolumn{7}{l}{10 - \cite{2007AN....328..928Z}; 11 - \cite{2008A+A...477..615S}; 12 - \cite{1986A+A...167..287C};} \\
\multicolumn{7}{l}{13 - Wolf, M.\ (2009, private communication); 14 - \cite{1999AJ....118.1831T};} \\
\multicolumn{7}{l}{15 - \cite{1985A+A...145..206A}; 16 - \cite{1984A+A...134..147A}; 17 - \cite{1995A+A...302..741C};} \\
\multicolumn{7}{l}{18 - \cite{2009A+A...498..821W}; 19 - \cite{2009NewA...14..413C}; 20 - \cite{1991A+A...246..397C};} \\
\multicolumn{7}{l}{21 - \cite{2005Ap+SS.296..141D}; 22 - \cite{2001AstL...27..712K}; 23 - \cite{v636cen};} \\
\multicolumn{7}{l}{24 - \cite{cmdra}.} \\
\endlastfoot
                                                                                
2  &    EM Car       &   0.0120$\pm$0.0005  &    0.081$\pm$0.005   &   $-$2.31$\pm$0.07\phs  &     & 1 \\ 
4  &    V478 Cyg     &   0.0158\pppp0.0007  &   0.1047\pppp0.0010  &   $-$2.26\pppp0.03\phs  &     & 2 \\ 
6  &    V578 Mon     &   0.0867\pppp0.0009  &       \nodata        &      \nodata            &  0  &                              \\
7  &    V453 Cyg     &    0.022\pppp0.002   &    0.054\pppp0.003   &   $-$2.28\pppp0.04\phs  &     & 3 \\ 
8  &    CW Cep       &   0.0293\pppp0.0006  &   0.0582\pppp0.0005  &   $-$2.10\pppp0.05\phs  &     & 2 \\ 
10 &    QX Car       &    0.278\pppp0.003   &  0.01222\pppp0.00022 &   $-$2.15\pppp0.05\phs  &     & 4 \\ 
12 &    V539 Ara     &    0.053\pppp0.001   &   0.0193\pppp0.0010  &   $-$2.32\pppp0.06\phs  &     & 5 \\ 
14 &    AG Per       &   0.0709\pppp0.0006  &   0.0262\pppp0.0004  &   $-$2.14\pppp0.06\phs  &     & 2 \\
\end{longtable}
\end{center}
}




{\small
\begin{center}
\begin{landscape}
\begin{longtable}{r@{\thinspace\thinspace\thinspace}l@{~~~}c@{~~}c@{~~}c@{~~}c@{~~}lc@{~~}c@{~~}c@{~~}c@{~~}l}

\caption{(continued)} \\
\hline\noalign{\smallskip}
 &        & $V$ (mag)  & $\pi_{\rm orb}$ (mas) & Dist (pc)
          &      &     & Mass                  & $\log L$      & $T_{\rm eff}$ & Radius        & \\
 & System & $P$ (d)    & $a$ (mas)             & [Fe/H] 
          & Star & SpT & (M$_{\odot}$)         & (L$_{\odot}$) & (K)           & (R$_{\odot}$) &  Ref \\
\noalign{\smallskip}\hline\noalign{\smallskip}
\endhead

\noalign{\smallskip}\hline\noalign{\smallskip\smallskip\smallskip}
\endfoot

\noalign{\smallskip}\hline\noalign{\smallskip}
\multicolumn{12}{l}{References: 1 - \cite{1998AJ....116.2536H}; 2 - \cite{2009ApJ...700.1349T}; 3 - \cite{2001AJ....121.1623H}; 4 - \cite{2007ApJ...659..626Z};
5 - \cite{2009ApJ...695.1527F};} \\
\multicolumn{12}{l}{6 - \cite{2006A+A...446..723M}; 7 - \cite{2008AJ....135..766M}; 8 - \cite{2005ApJ...627..464B}; 9 - \cite{2006AJ....131.2695B};
10 - \cite{1999ApJ...515..356B};} \\
\multicolumn{12}{l}{11 - \cite{1999ApJ...527..360B}; 12 - \cite{2005AJ....130.2866M}; 13 - \cite{1997AJ....113.1095F}; 14 - \cite{2006ApJ...644.1193B};
15 - \cite{1987AJ.....93.1236T};} \\
\multicolumn{12}{l}{16 - \cite{2002AJ....124.1716T}; 17 - \cite{2007A+A...464..635B}; 18 - \cite{1999A+A...351..619F}; 19 - \cite{2000A+A...364..665S};
20 - \cite{2001AJ....121.1607B};} \\
\multicolumn{12}{l}{21 - \cite{2000AJ....120.1106B}.} 
\endlastfoot

\caption{Parameters for 23 astrometric binary systems.\label{tableMonly}} \\
\hline\noalign{\smallskip}
 &        & $V$ (mag)  & $\pi_{\rm orb}$ (mas) & Dist (pc)
          &      &     & Mass                  & $\log L$      & $T_{\rm eff}$ & Radius        & \\
 & System & $P$ (d)    & $a$ (mas)             & [Fe/H] 
          & Star & SpT & (M$_{\odot}$)         & (L$_{\odot}$) & (K)           & (R$_{\odot}$) &  Ref \\
\noalign{\smallskip}\hline\noalign{\smallskip}
\endfirsthead

 1 & $\zeta^1$ UMa   &   2.27 &  39.4$\pm$0.3\phn     & 25.38$\pm$0.19\phn   &  A  & A2V    &    2.50$\pm$0.07    &      1.522$\pm$0.027   & 9000$\pm$200\phn & 2.38$\pm$0.13      &  1 \\ 
   & HD 116656       &  20.54 &    9.83\pppp0.03      &     \nodata          &  B  & A2V    &    2.43\pppp0.07    &      1.522\pppp0.027   & 9000\pppp200\phn & 2.38\pppp0.13      & \\
\noalign{\smallskip}  
 2 & $\alpha$ Aur    &   0.07 & 76.67\pppp0.17\phn    & 13.042\pppp0.028\phn &  A  & G8III  &   2.466\pppp0.018   &      1.900\pppp0.026   & 4920\pppp70\pht  & 11.87\pppp0.56\phn &   2 \\ 
   & HD 34029        & 104.02 & 56.408\pppp0.025\phn  & $-$0.34\pppp0.07\phs &  B  & G0III  &   2.443\pppp0.013   &      1.858\pppp0.022   & 5680\pppp70\pht  & 8.75\pppp0.32      & \\
\noalign{\smallskip}  
 3 & $o$ Leo         &   3.52 &  24.2\pppp0.1\phn     &    41.4\pppp0.1\phn  &  A  & F9III  &    2.12\pppp0.01    &      1.595\pppp0.027   & 6000\pppp200\phn & 5.8\pppp0.4        &  3 \\ 
   & HD 83808/9      &  14.50 &    4.46\pppp0.01      &     \nodata          &  B  & A5m    &    1.87\pppp0.01    &      1.188\pppp0.028   & 7600\pppp400\phn & 2.3\pppp0.3        & \\
\noalign{\smallskip}  
 4 & $\lambda$ Vir   &   4.52 & 18.81\pppp0.10\phn    &   53.16\pppp0.29\phn &  A  & A1m    &   1.897\pppp0.016   &      1.319\pppp0.005   & 8280\pppp200\phn & 2.22\pppp0.11      &  4 \\ 
   & HD 125337       & 206.73 & 19.759\pppp0.079\phn  &   \nodata            &  B  & A1m    &   1.721\pppp0.023   &      1.100\pppp0.005   & 8280\pppp200\phn & 1.73\pppp0.08      & \\
\end{longtable}
\end{landscape}
\end{center}
}

\end{document}